\documentclass[11pt,preprint]{aastex}
\usepackage{times}
\usepackage{multirow}

\def\ltsima{$\; \buildrel < \over \sim \;$}
\def\simlt{\lower.5ex\hbox{\ltsima}}
\def\gtsima{$\; \buildrel > \over \sim \;$}
\def\simgt{\lower.5ex\hbox{\gtsima}}

\begin{document}
\title{Where are the Fossils of the First Galaxies? I. Local Volume Maps and Properties of the Undetected Dwarfs}
\author{Mia S. Bovill and Massimo Ricotti}
\affil{Department of Astronomy, University of Maryland, College Park,
  MD 20742}
\email{msbovill@astro.umd.edu,ricotti@astro.umd.edu}

\begin{abstract}
  We present a new method for generating initial conditions for $\Lambda$CDM
  N-body simulations which provides the dynamical range necessary to
  follow the evolution and distribution of the fossils of the first
  galaxies on Local Volume, $5-10$~Mpc, scales. The initial
  distribution of particles represents the position, velocity and mass
  distribution of the dark and luminous halos extracted from
  pre-reionization simulations. We confirm previous results that
  ultra-faint dwarfs have properties compatible with being well
  preserved fossils of the first galaxies. However, because the
  brightest pre-reionization dwarfs form preferentially in biased
  regions, they most likely merge into non-fossil halos with circular
  velocities $>20-30$~km/s. Hence, we find that the maximum luminosity
  of true-fossils in the Milky Way is $L_V<10^5$~L$_\odot$, casting
  doubts on the interpretation that some classical dSphs are
  true-fossils. In addition, we argue that most ultra-faints at small
  galactocentric distance, $R<50$~kpc, had their stellar properties
  modified by tides, while a large population of fossils is still
  undetected due to their extremely low surface brightness
  $\log(\Sigma_V) < -1.4$. We estimate that the region outside
  $R_{50}$ ($\sim 400$~kpc) up to $1$~Mpc from the Milky Way contains
  about a hundred true fossils of the first galaxies with V-band
  luminosity $10^3 - 10^5$~L$_\odot$ and half-light radii, $r_{hl} \sim
  100-1000$~pc.
\end{abstract}

\section{Introduction}

Simulations of the formation of the first galaxies have matured,
however there are few observational constraints on the models. The
overabundance of dark matter satellites near the Milky Way when
compared to observations of luminous satellites (missing galactic
satellites), and lack of dwarfs in the voids (void phenomenon) suggest
that these early galaxies may be too faint to be detected directly at
high-redshift even for JWST \citep{RicottiGnedinShull:08,Johnsonetal:09}. However, we {\it{can}} detect their fossil remnants
in the local universe. Hierarchical formation scenarios predict these
first galaxies formed before reionization in dark matter minihalos
with masses $ \simlt 10^8 M_\odot$.  Those that survive to the present
constitute, in part, a sub-population of satellites around larger
halos. Recent observational and theoretical advances allow us to
compare simulated primordial galaxies to the observations of the
faintest galaxies in the Local Group \citep{Belokurovetal:06a, Belokurovetal:07, Irwinetal:07, Walshetal:07, Willmanetal:05AJ, Willmanetal:05ApJ,Zuckeretal:06b, Zuckeretal:06a, Gehaetal:09} and
constrain models of star formation in the early universe.

The formation of the first galaxies before reionization is regulated
by complex feedback effects acting on cosmological distance
scales. These self-regulation mechanisms have dramatic effects on the
number and luminosity of the first galaxies. The gas in minihalos with
a circular velocity, $v_{max} < 20$~km~s$^{-1}$ is heated to $T \simlt
10,000$~K during virialization. At this temperature, a gas of
primordial composition is unable to cool and initiate star formation
unless it can form and retain a sufficient density of $H_2$. Although
$H_2$ is destroyed by dissociating UV radiation in the Lyman-Warner
bands, its formation can be catalyzed by hydrogen ionizing radiation,
via the formation of $H^-$ \citep{HaimanReesLoeb:96}. Thus radiative
transfer is necessary to simulate $H_2$ formation and destruction in
the optically thick early universe. Along with other relevant physics,
3D radiative transfer was included in Ricotti et al (2002a,b,2008),
hereafter referred to as pre-reionization simulations.

Ricotti et al.'s pre-reionization simulations produce a population of
primordial galaxies with stellar properties consistent with a subset
of the classical dSphs \citep{RicottiGnedin:05}(hereafter RG05) and a
majority of the recently discovered ultra-faint dwarfs
\citep{BovillRicotti:09,Ricotti:10}(hereafter BR09). Over the last
five years the ultra-faints doubled the census of Milky Way and M31
satellites \citep{Willman:10}(and references therein). The new Milky
Way dwarfs have V-band luminosities $< 10^5 L_\odot$, half-light radii
$r_{hl}\sim 20-300$~kpc, metallicities $[Fe/H] \simlt -2$, mass to
light ratios $\simgt 100 M_\odot/L_\odot$, stellar velocity
dispersions $\sim 2-10$~km~s$^{-1}$ and are $30 - 400$~kpc from the
Milky Way. The new M31 dwarfs have $L_V \simgt 10^5$~L$_\odot$ with
$r_{hl}$ systematically greater than their Milky Way
counterparts. Except for Leo T, both groups are dominated by an old
metal poor population and are devoid of gas. In addition, a subset of
the M31 dwarfs are falling into their host halo for the first time \citep{Majewskietal:07}. These new dwarfs provide the
observational laboratory we need to test our model of the properties
and distribution of the fossils of the first galaxies.

\begin{figure}
\centering
\plotone{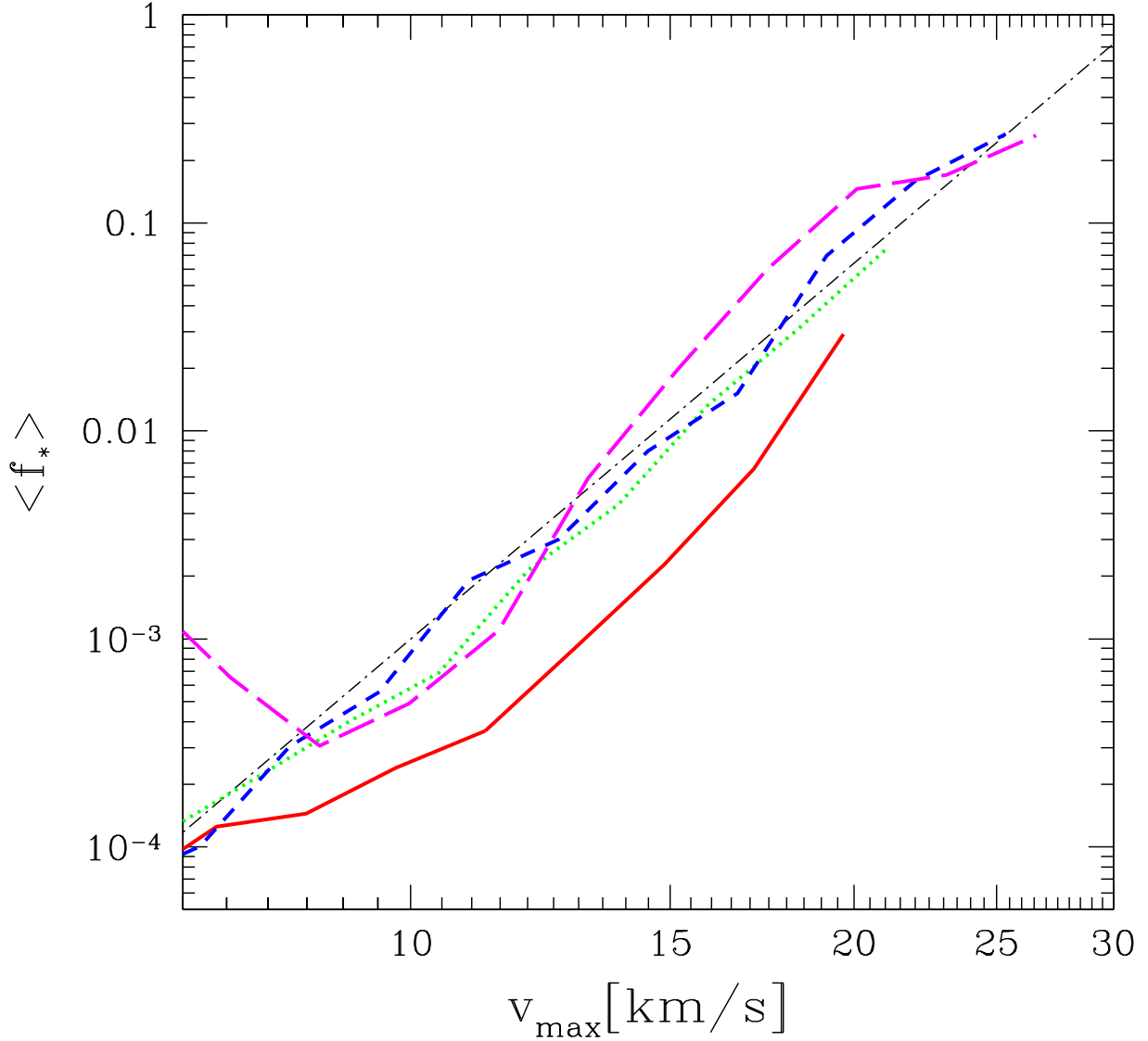}
\caption {The mean star formation efficiency $\langle f_* \rangle$ as a function of $v_{max}$ for all galaxies from RG05 at redshifts $z=14.6, 12.5, 10.2$ and $8.3$. Here, the circular velocity of the galaxies is $v_{max}=17 km/s (M/10^8 M_\odot)^{1/3}[(1+z)/10]^{1/2}$, calculated from their total mass $M$ and redshift $z$. The star formation efficiency is the fraction of baryons within the virial radius of the halo that is converted into stars $f_* \equiv (\Omega_m/\Omega_b)M_*/M$.}
\label{FIG.fstvmax}
\end{figure}

The large scatter in the properties of Ricotti et al. galaxies with the same mass follows from the physics governing star formation in minihalos. Figure~\ref{FIG.fstvmax} shows that in minihalos the mean star formation efficiency, $f_*$, is much smaller than the star formation efficiency per free fall time ($\epsilon_*=10\%$) that appears as a free parameter in the adopted sub-grid recipe for star formation: ${\dot \rho_*}=\epsilon_* \rho_g/t_{dyn}$. In contrast, in halos with mass $M > 10^8-10^9$~M$_\odot$ we find $f_* \simeq \epsilon_*$. The value of $f_*$ is smaller than $\epsilon_*$ because in small halos the gas available for star formation is reduced with respect to the mean cosmic value. Two effects are dominant in reducing the amount of gas available for star formation in small mass galactic halos: i) photo-heating of the IGM reduces the amount of gas that falls into the potential wells with respect to the mean cosmic value; ii) massive stars inside luminous galaxies ionize and heat their ISM expelling most of the gas available for further star formation (see \cite{RicottiGnedinShull:08}). SN explosions also contribute to produce galactic winds but UV radiation from massive stars operates on a shorter time scale and is most effective in clearing out the majority of the gas in the halo.  Thus, the number of massive stars that can be produced in such small mass galaxies is regulated by feedback loops.  As a result $\langle f_* \rangle$ is much smaller than $\epsilon_*$ and its value is rather insensitive to the assumed value of $\epsilon_*$. 

Although, as shown in Figure~\ref{FIG.fstvmax}, the mean star formation efficiency in the first galaxies is roughly a power law $\langle f_* \rangle \propto M^2 \propto v_{cir}^6$ , the scatter about this relationship is large, especially for the smallest mass halos in the plot \citep{RicottiGnedinShull:08}. Halos with masses $M<10^7$~M$_\odot$ can be either luminous or dark, depending on the environment in which they reside.  For a given mass, halos that are relatively isolated and far from luminous galaxies tend to be the least luminous or completely dark. The proximity to a source of ionizing radiation and metals has a positive effect on triggering star formation in dark halos.  Ionizing radiation stimulates molecular hydrogen formation and enhance the cooling rate of the gas \citep{RicottiGnedinShull:01,RicottiGnedinShull:02a}. Similarly metal pollution stimulates gas cooling. Hence, the first galaxies are highly biased. The most luminous of them are most likely to merge into more massive objects and do not survive to the present.  This explains why in the simulations shown in the present study, contrary to our naive initial expectation, we did not find fossil dwarfs with $L_V>10^6$~L$_\odot$ around the Milky Way.

\cite{GnedinKravtsov:06} (hereafter GK06) uses this approximation in conjunction with high-resolution N-body simulations of the Local Group, to evolve a population of dwarf galaxies around a Milky Way mass halo from $z =70$ to $z = 0$.  For details of the simulations, see \S~2 in GK06. GK06 defines a fossil as a simulated halo which survives to $z = 0$ and remains below the critical circular velocity of $v_{filt} = 30$~km~s$^{-1}$ with no appreciable tidal stripping. They calculate the probability, $P_S(v_{max}, r)$, of a luminous halo with a given maximum circular velocity $v_{max}$ to survive from $z = 8$ (the final redshift of the RG05 simulation) to $z = 0$.  For a given $v_{max}$, the number of surviving dwarfs at $z = 0$ is $N(v_{max}, z = 8)P_S(v_{max},r)$, where $P_S$ is the survival probability for a satellite at a distance $r$ from the host halo. The surviving halos are assigned a luminosity based on the $L_V$ versus $v_{max}$ relationship from RG05. At $z = 0$, GK06 has a population of dwarf galaxies with a resolution limit of $v_{max} = 13$~km~s$^{-1}$. The halos are statistically assigned luminosities from the $L_V-v_{max}$ relation given in Figure 3 of GK06. Unfortunately, with this relation, the $13$~km~s$^{-1}$ corresponds to a lower luminosity limit of $L_V \sim 10^5$~L$_{\odot}$, which includes Leo T and Canes Venatici I, but excludes all the other new ultra-faint Milky Way satellites.

In this paper, we describe and test a novel method of generating N-body initial conditions which allows us to follow the evolution, merger rates and tidal destruction of pre-reionization halos to present day and to overcome some of the limitations of the GK06 method.  The initial distribution of particles in the N-body simulations represents the position, velocity and mass distribution of the dark and luminous halos extracted from pre-reionization simulations.  Our simulations have a sufficiently large volume and dynamical range to explore the distribution of fossil galaxies outside the Local Group, in nearby filaments and voids using limited computational resources.

Our method improves on the GK06 work by removing the constraints that preclude a comparison of the GK06 simulations with the observed distributions of the ultra faints: (1) Due to the resolution of their N-body simulations, GK06 cannot resolve dwarfs with circular velocity, $v_{max}$, $<13$~km~s$^{-1}$ which roughly corresponds to a simulated dwarf with $L_V<10^5 L_\odot$. With two exceptions, no ultra faint dwarfs have $L_V>10^5 L_\odot$ (CVn I \citep{Zuckeretal:06a} and Leo T \citep{deJongetal:08}). (2) The statistical matching of the baryonic properties of the pre-reionization halos to equivalent $z=0$ halos in their N-body simulation, does not allow GK06 to account for mergers of pre-reionization halos after reionization. While the majority of mergers would not involve two luminous pre-reionization halos, the effect cannot be ruled out {\it{apriory}}. (3) The GK06 statistical matching also does not account for the clustering bias of the most luminous pre-reionization halos. The formation efficiency of $H_2$ is dependent on stochastic effects, so the most luminous pre-reionization halos form in the highest density regions of the \cite{RicottiGnedinShull:02b} simulations and are more likely to have undergone a merger with another massive, luminous pre-reionization halo. (4) Extracting the baryonic properties of their fossils at $z=0$ from the final output of the pre-reionization simulation does not allow GK06 to account for cosmic variance. By $z=0$, the faster evolution of structure in over-dense regions (ie. Local Group) and the slower structural evolution of under-dense regions (ie. Local Void) have produced significant variance in the numbers and types of objects seen in both.

The paper is organized as follows. In \S~\ref{SEC.NM} we describe our initial conditions in detail before comparing our results to traditional CDM N-body initial conditions in \S~\ref{SEC.tests}. We show that with this method we can easily achieve the resolution necessary to study the distribution of ultra-faint dwarfs at $z=0$ in a volume similar to the Local Volume.  Initial results are in \S~\ref{SEC.Res} and the remainder are presented in \cite{BovillRicotti:10b}, Paper II of this series. Our simulations are compared to the GK06 results in \S~\ref{SEC.gk06} in which we propose a maximum luminosity threshold for primordial fossils. Finally we compare the properties of our simulated primordial dwarfs to observations of Local Group dwarfs in \S~\ref{SEC.prop}, and present observational tests for the primordial formation model in \S~\ref{SEC.obser}. A summary and conclusions are presented in \S~\ref{summary}.

\section{Numerical Method} 
\label{SEC.NM}

To achieve the resolution necessary to study the ultra-faint dwarfs in a $z=0$ volume equivalent to the Local Volume, we developed a method for generating initial conditions for N-body simulations which provides the required mass resolution, while using only limited computational resources. Our simulations allow us to trace the merger rate and tidal stripping of the first galaxies from reionization to the modern epoch. Traditional initial conditions for CDM simulations begin with an evenly distributed grid of uniform mass particles before the positions and velocities are perturbed according to a given power spectrum. Our method follows the same concept, except the initial distribution of the particles is not a uniform grid, but represents the distribution of halos in the final outputs of a $1$Mpc$^{3}$ high resolution cosmological hydrodynamical simulation run to $z=8.3$ \citep{RicottiGnedinShull:02b} (hereafter we refer to these as the pre-reionization simulations and the halos found in their $1$~Mpc$^{3}$ outputs as pre-reionization halos). Thus, each particle represents a dark or luminous halo with a different mass and given stellar properties.

All of the initial conditions described in this section were run from their initial redshift, $z_{init}$, to $z=0$ using Gadget 2 \citep{Springel:05} on the Maryland HPCC Deepthought and analyzed with the halo finder AHF \citep{KnollmannKnebe:09}. 

To construct our high resolution region, we produce a lattice of the pre-reionization simulation $z=8.3$ output. This gives us a grid similar to that used in traditional CDM, except power on scales below 1 Mpc is already present through the positions of the pre-reionzation halos. To add the larger scale power, we perturb the particle positions and velocities of the pre-reionization halos according to a power spectrum with no power for modes $l < 1$~Mpc. This method is similar to the one described in \cite{TormenBertschinger:96} (see Appendix~\ref{FO}  for details). The high resolution region $\sim10$~Mpc on a side, with a mass resolution of $\sim3.2 \times 10^5 M_\odot$, is embedded in a coarse resolution volume 50 Mpcs on a side containing $250^{3}$ particles at $z=8.3$.

When compared to traditional zoom simulations, our high resolution region has several key differences. Primarily, each of our particles represents a resolved halo from the pre-reionziation simulations. Each of these pre-reionization halos has a set of dark matter and stellar properties derived at $z=8.3$. This technique allows us to push our simulations to higher mass resolutions over a `Local Volume' sized region without a prohibitive increase in the number of particles. However, this technique precludes us from determining detailed density profiles of pre-reionization halos at $z=0$. The stellar properties of the pre-reionzation halos are preserved through the unique IDs of each particle in our simulation.  If, in the modern epoch, a given pre-reionization halo is in a halo whose maximum circular velocity has never exceeded the filtering velocity it has not accreted gas from the IGM after reionization. This filtering velocity, $v_{filter}$ represents the critical value for the maximum circular velocity of the halo below which star formation is suppressed by the reheating of the IGM via reionization feedback \citep{Gnedin:00c,BabulRees:92,Efstathiou:92,Shapiroetal:94,Shapiroetal:04,ThoulWeinburg:96,Quinnetal:96,NavarroSteinmetz:97,SusaUmemura:04,Bensonetal:06,Hoeftetal:06}.  The subsequent lack of star formation in these low mass halos allows us to approximate its present day observable properties from those at reionization. We consider the initial conditions built using the method described above as our first order simulations, specifically, runs A, B and C (Table~\ref{TAB.runs}). The initial conditions for run D, which are significantly different than those described for runs A-C, are described in the next section. A detailed methodology for both sets of initial conditions is given in the appendix. 

For the remainder of this work, we focus on Run C since Runs A and B do not have the resolution necessary to study the dwarf populations inside the Milky Way halo.

\subsection{Approximating Cosmic Variance}

For our first order simulations (see Table~\ref{TAB.runs}), we assume that every part of our `Local Volume' evolves at the rate associated with the mean density of the universe, before and after reionization. However, there are deviations from this mean due to linear perturbations on large ($> 1$~Mpc) scales . The evolution of a given region depends on its mean density with regions of higher density evolving faster than their lower density counterparts \citep{Cole:97,Reedetal:07,Crainetal:09}. As a result, halos will collapse, and form stars, at later times in the voids compared to the filaments. To account for this effect, we relate the over-density or under-density of each region of our high resolution region to the speed of its evolution. We express the evolution of a region as a function of its densities as $z_{eff}=z_{init}+\Delta z$, where $z_{eff}$ is the effective redshift, $z_{init}$ is the redshift of the simulations and the effective redshift of a region whose local density is the average density of the universe, $\rho_o(z_o$), and:
\begin{equation}
\Delta z = (1+z_{init})[(1+\delta)^{-0.6}-1)]
\end{equation}
is the correction to $z_{init}$ due to $\delta$, the local over-density or under-density of a given region.

To approximate this variance effect, we produce a set second order initial conditions as in \cite{Cole:97} (run D). The primary difference between runs D and C lies in the construction of the high resolution volume. Instead of using a single pre-reionization simulation output at $z=8.3$ (runs A-C), we use outputs at multiple redshifts ($z=8.3-14$) to approximate the differential evolution of the universe up to $z_{init}=10.2$. Before constructing our high resolution region, we calculate the effective redshift of each $1$~Mpc$^{3}$ sub-volume. Each sub-volume is then assigned a pre-reionization output based on its effective redshift, with the lowest density voids at $z_{eff}=14$ and highest density regions at $z_{eff}=8.3$. For additional details, see Appendix~\ref{SO}. 

In addition to accounting for cosmic variance, comparisons of runs A-C and run D allow us to probe two different reionization scenarios. Since we cannot account for baryonic evolution after ``reionization'' when the pre-reionization outputs are transformed into our N-body simulation, we assume that the photo-evaportation and reheating during reionization precludes any further baryonic evolution in the minihalos. During reionization by UV, we assume the IGM throughout our volume was reheated to $\sim10^4$~K \citep{RicottiOstriker:04}. We also assume that the entire volume was reionized at $\sim z_{init}$. For runs A-C this approximates reheating at $z_{init}\sim 8.3$ by UV photons generated by stars in the first galaxies \citep{Sokasianetal:04,WiseC:09}. Since the voids evolve at a slower rate than the filaments, using the same pre-reionization output for our entire simulation is effectively allowing the low density regions to evolve for a longer time before their IGM is reheated to 10$^4$ K. This is consistent with reionization and reheating beginning in the filaments before spreading into the voids. Since low mass halos in the voids would have had more time to accrete gas and form stars before the reheating cut off their gas supply, we expect the voids in runs A-C to be significantly brighter than those in run D (Figure~\ref{LSS})

Since each $1$~Mpc$^3$ subvolume in Run D used a pre-reionization output consistent with its effective redshift, the entire high resolution region has been given the same amount of time to evolve. When we transition from the pre-reionization output to our N-body simulations, Run D does not allow low mass halos in the low density regions to continue to evolve as the denser filaments are reheated. Instead, Run D approximates a universe in which all of space is reionized and reheated at approximately the same time. Uniform reheating of the filaments and voids is a characteristic of reionization and reheating from X-rays produced by remnants of the first stars accreting from the ISM at higher redshift \citep{Venkatesanetal:01, RicottiOstriker:04,RicottiOstrikerGnedin:05,Ripamontietal:08,ShullV:08}. X-rays could also be produced by primordial black hole binaries \citep{Mirabeletal:11,Saigoetal:04}. When compared to reheating by UV radiation, X-ray reheating produces noticeably darker voids.

\subsection{A Note on the Halo Occupation Distribution}

The luminosities of the $z=0$ halos are determined by the luminosities of their component pre-reionization halos. These pre-reionization luminosities are taken directly from the Ricotti et al (2002a,b) pre-reionization simulations and are determined by the feedback prescriptions and star formation efficiencies used in that work. Predictions made based on the resulting luminosity function and galactocentric radial distribution are a result of the primordial formation model we assume for the smallest dSphs. Note, that the match of luminosity and dark matter mass in this simulation is not statistical, but rather a direct result the distribution of the remnants of the first galaxies in the modern epoch.

\begin{table}
\centering
\begin{tabular}{ c c c c c c c }
\hline
Name & IC Method & Volume & HR Volume & Mass Res. & $\epsilon$ & $z_{init}$ \\
& & (Mpc$^{3}$) & (Mpc$^{3}$) & ($10^6 M_\odot$) & (kpc) & \\
\hline
\hline
A & 1$^{st}$ order & $50^3$ & $\sim 9^3$ & 3.16 & 1 & 8.3 \\
B & 1$^{st}$ order & $50^3$ & $\sim 9^3$ & 1.0 & 1 & 8.3 \\
C & 1$^{st}$ order & $50^3$ & $\sim 9^3$ & 0.316 & 1 & 8.3 \\
D & 2$^{nd}$ order & $50^3$ & $\sim 9^3$ & 0.316 & 1 & 10.2 \\
\hline
\end{tabular}
\caption{Table of simulation runs. From left to right the columns are (1) run identifier, (2) type of initial conditions, and approximate reionization model, (3) size of low resolution volume in Mpc$^3$, (4) approximate size of the high resolution volume in Mpc$^3$, (5) mass of dark tracer particles in $10^6 M_\odot$, (6) softening length in kpc, and (7) initial redshift of the zoom simulation.}
\label{TAB.runs}
\end{table}

\section{Tests of the Method}
\label{SEC.tests}

In this section, we present consistency checks of our method to confirm that it reproduces known results from previous CDM simulations. We also discuss numerical effects introduced by our use of a spectrum of masses in our high resolution region. First, we confirm that the large scale structure and clustering of matter is consistent with traditional CDM simulations run with constant mass per particle.  Then, we see that the halo mass function is consistent with the mass function of halos derived from the Press-Schechter formalism \citep{PressSchechter:74}. Finally, we confirm that the number of subhalos and their Galactocentric distribution agree with the published results of the Via Lactea (Diemand et al. 2008) and Aquarius (Springel et al. 2008) simulations and that mass loss due to tidal stripping of the z = 0 halos is also in agreement with Kravtsov et al. (2004).

\begin{figure}
\centering
\includegraphics[width=84mm,height=68mm]{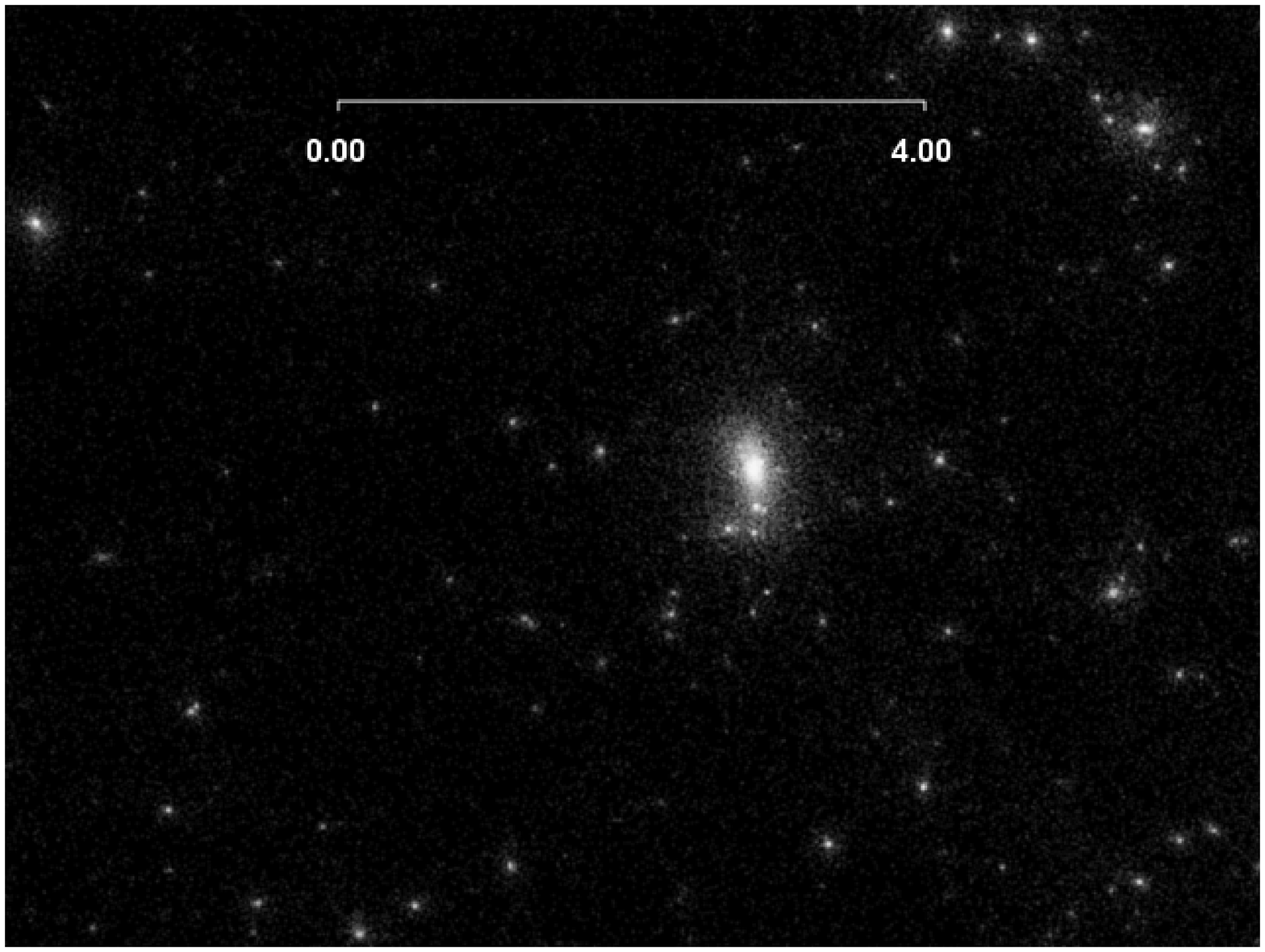}
\includegraphics[width=84mm,height=68mm]{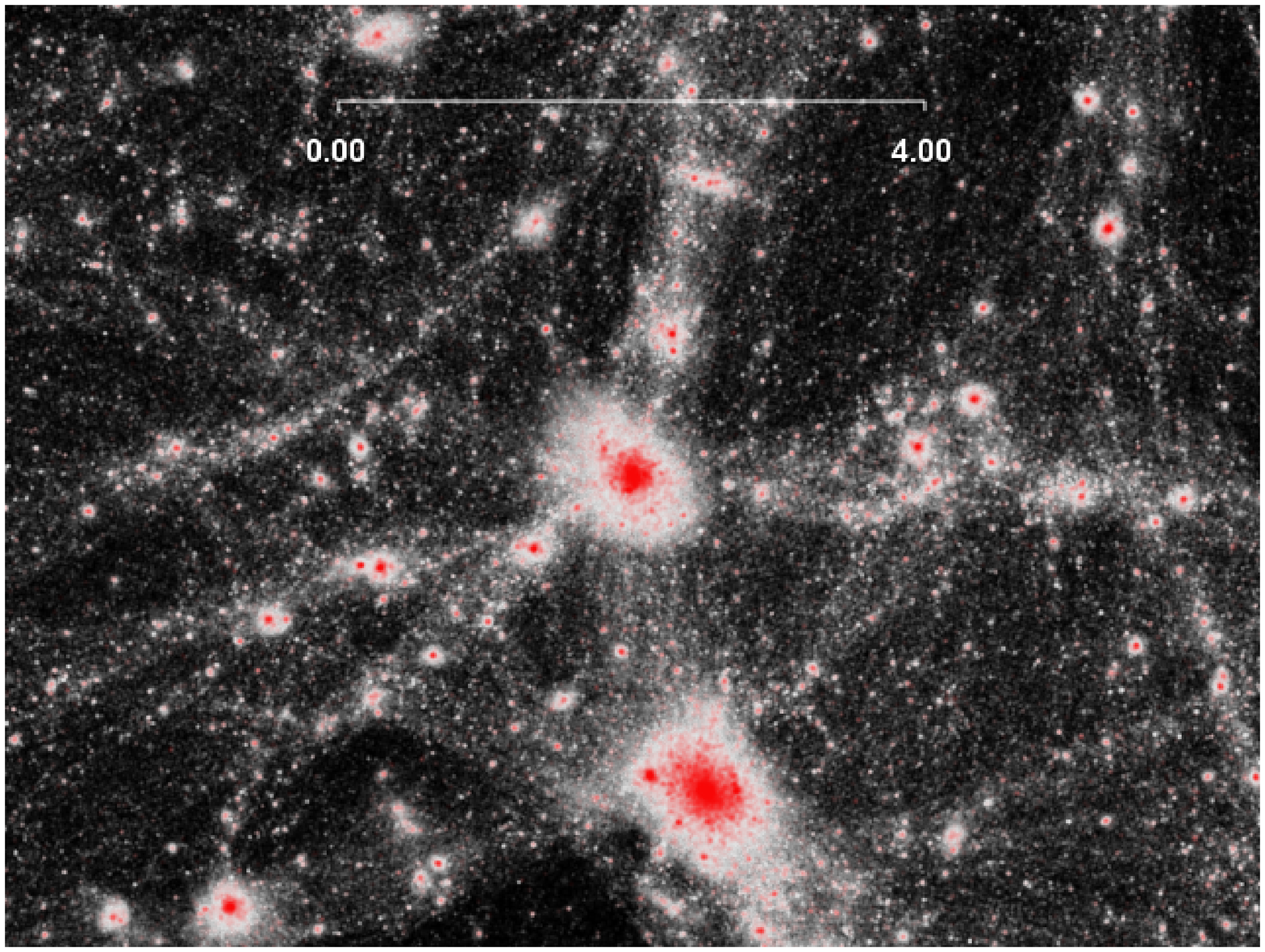}
\includegraphics[width=84mm,height=68mm]{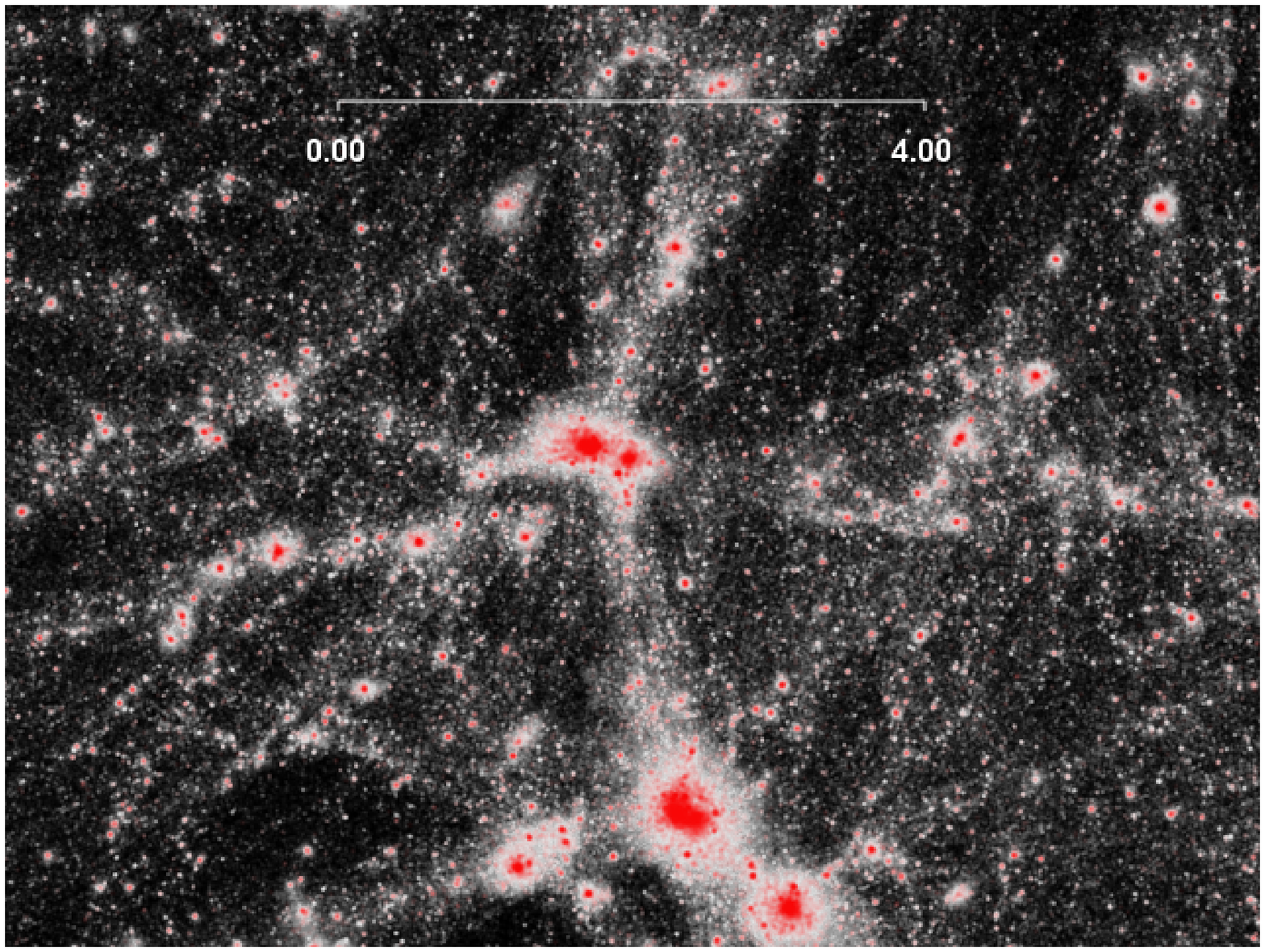}
\caption{Large scale structure of the same region of our simulations for the low resolution simulations (top), Run C (middle) and Run D (bottom). For Runs C and D white shows the halos with no luminous component while the larger, red points show the luminous pre-reionization halos.  The color of the latter does not depend on luminosity. The bar across the top of each panel shows the scale in Mpc.}
\label{LSS}
\end{figure}

Figure~\ref{LSS} shows a region of our Local Volume $5$~Mpc across at $z=0$. From top to bottom are the coarse simulation, Run C and Run D seen from the same viewing angle. For Runs C and D, the luminous pre-reionization halos are shown as large red dots plotted over the white distribution of dark tracer particles.  We find both Run C and D reproduce the large scale structure seen in the coarse resolution simulation.

\subsection{Mass Resolution}
\label{SEC.mass}

Our simulations produce maps of the present day distribution and properties of pre-reionization fossils in a $5^3$~Mpc$^3$ volume around a Milky Way type halo and in local filaments and voids. One of our goals is to map the distribution and properties of fossil galaxies outside the large hosts. This is done to quantify the number and properties of luminous dwarfs in the voids if stars formed in minihalos before reionization. These dwarfs would have evolved in relative isolation, and, if found by observations, would represent unambiguous and unperturbed fossils of the first galaxies. However, at this time, the only observational sample to which we can compare our simulations is the classical dSphs and ultra-faint dwarfs near the Milky Way and M31. The faintest known dwarfs ($L_V<10^3 L_\odot$) are found at less than $50$~kpc from the Galactic center. Observations are likely incomplete at $R>50$~kpc (approximately one quarter the virial radius) \citep{Koposovetal:07,Walshetal:09}. To compare our simulations to observations of the faintest known dwarfs, we must resolve halos within 100 kpc of the Milky Way center.

\begin{figure*}
\centering
\plottwo{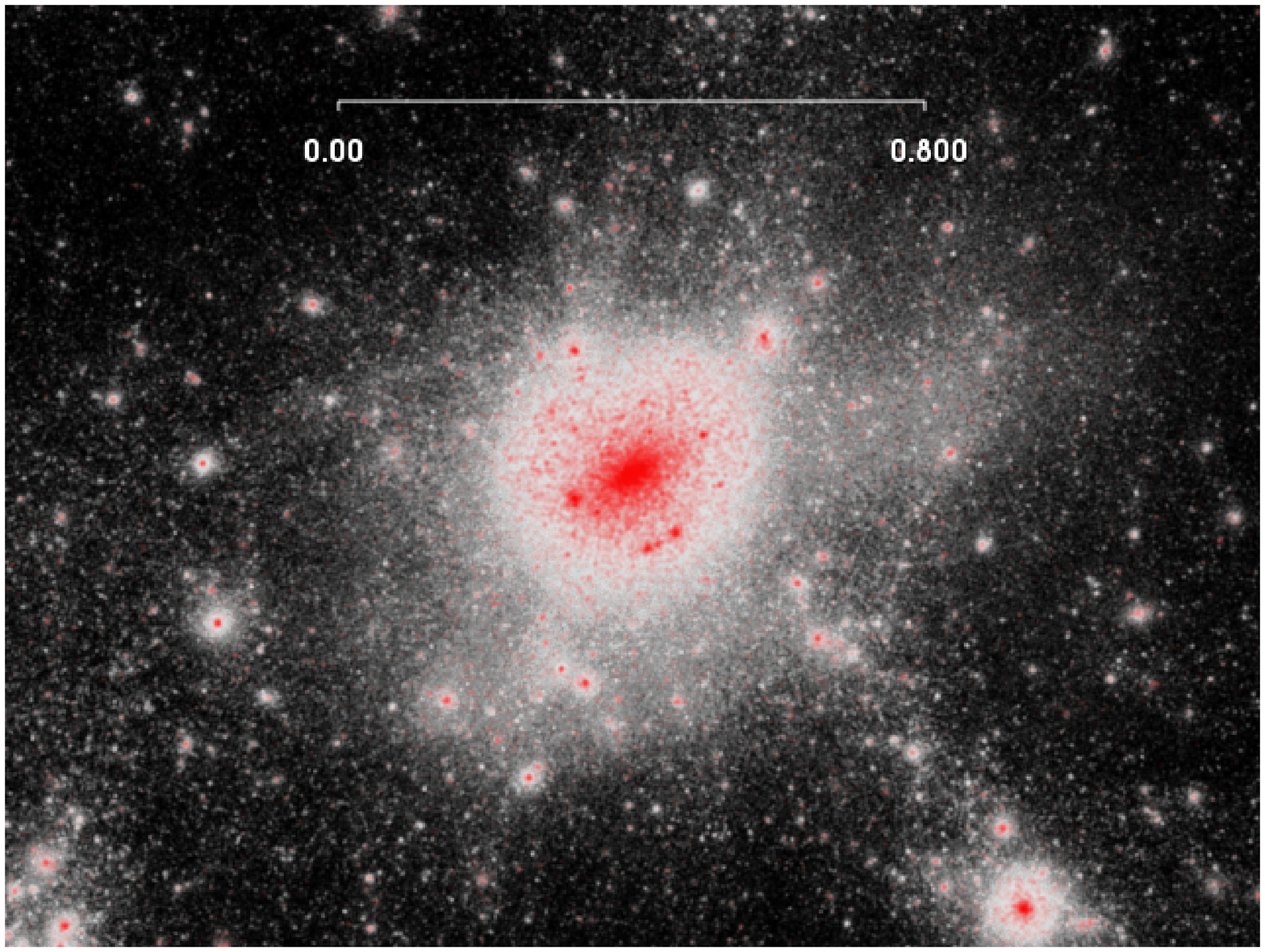}{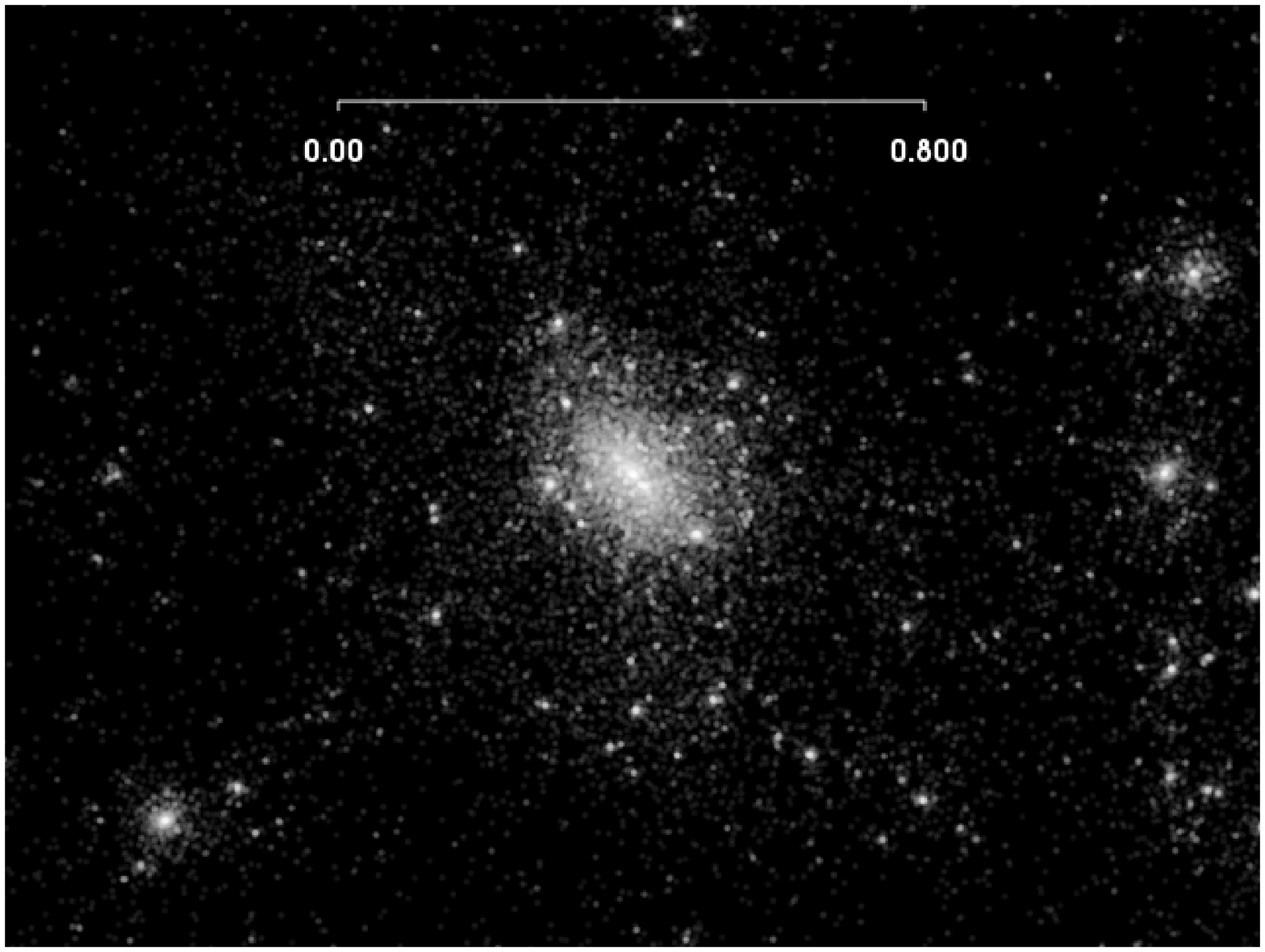}
\plottwo{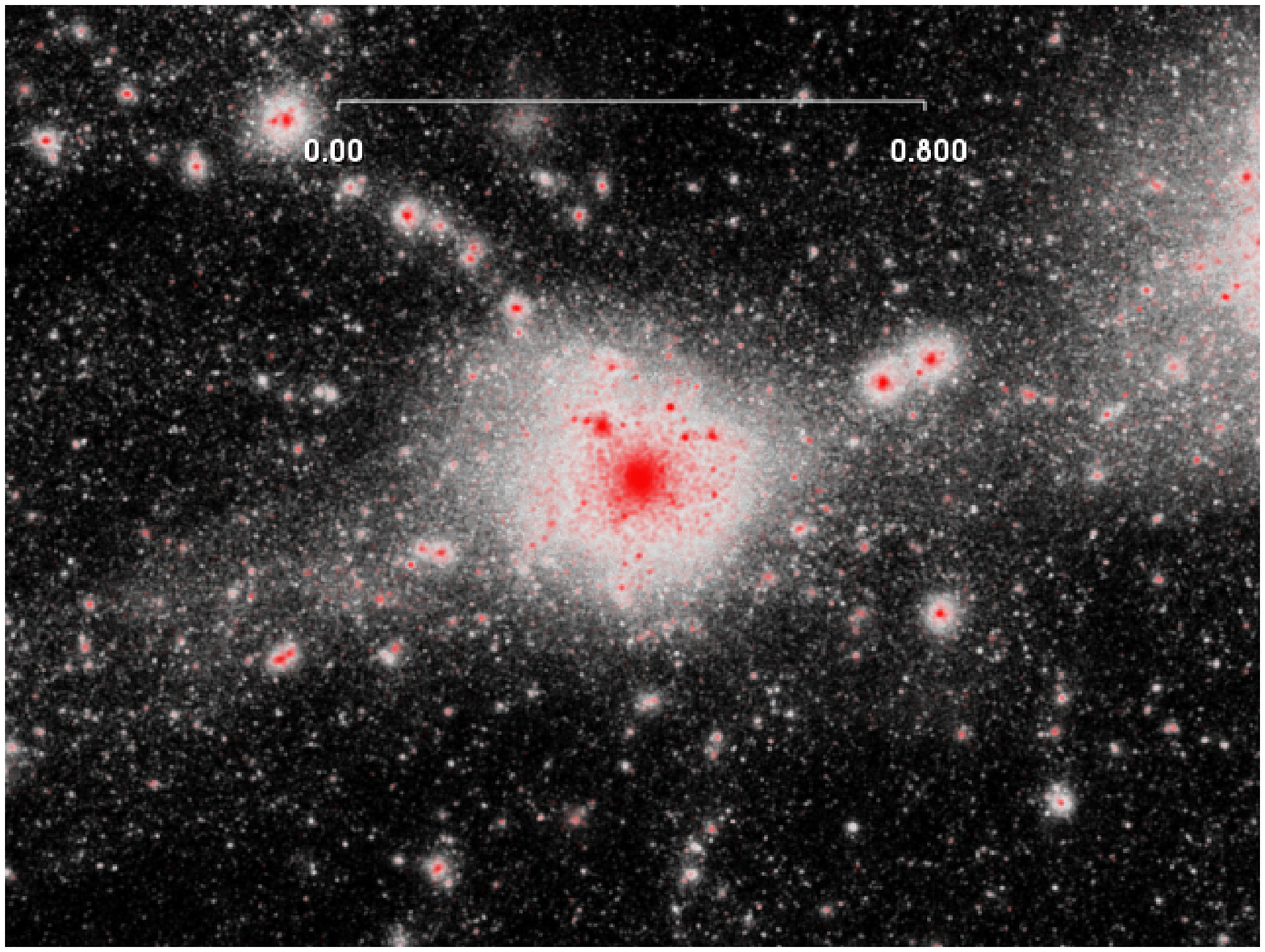}{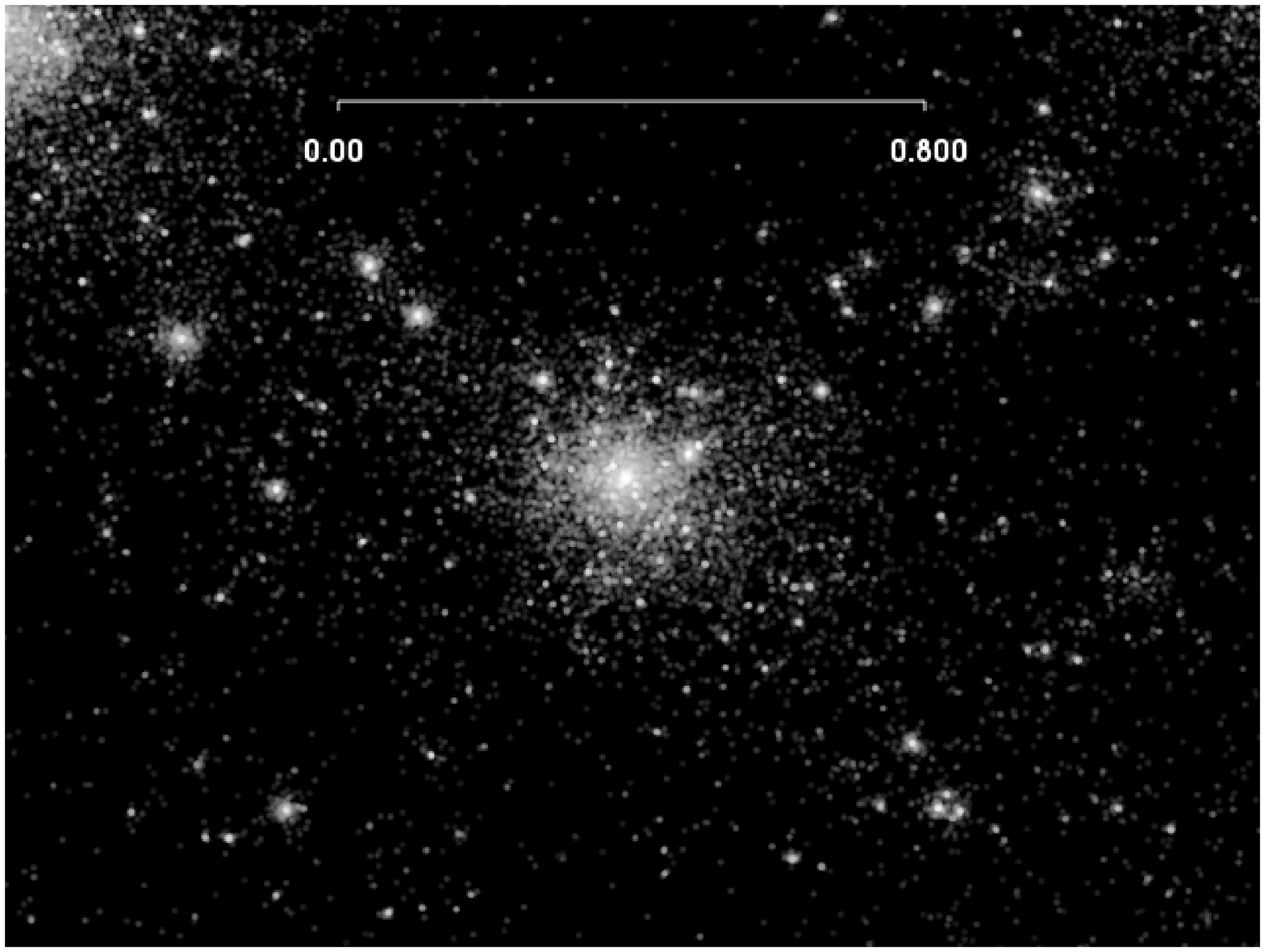}
\plottwo{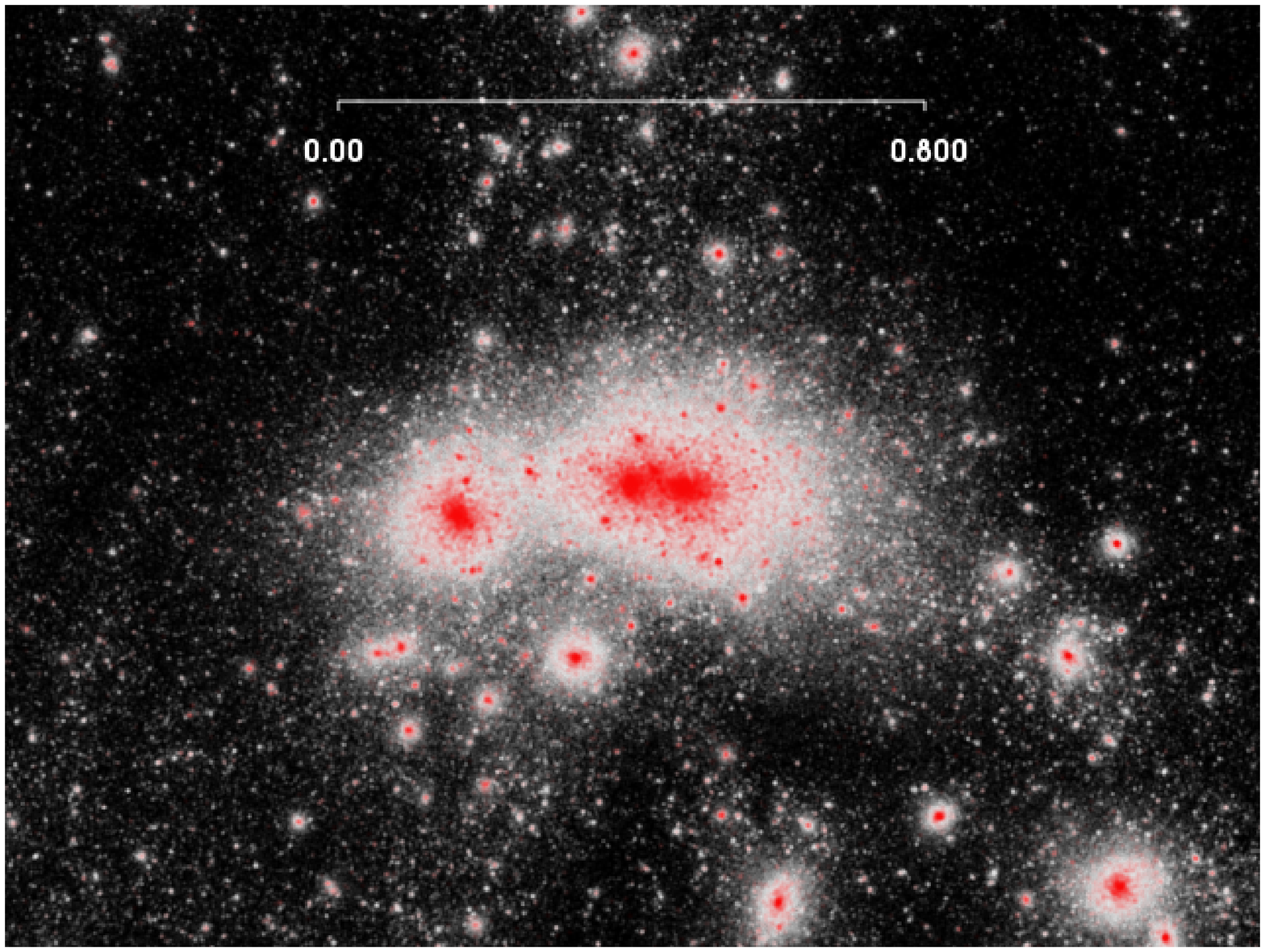}{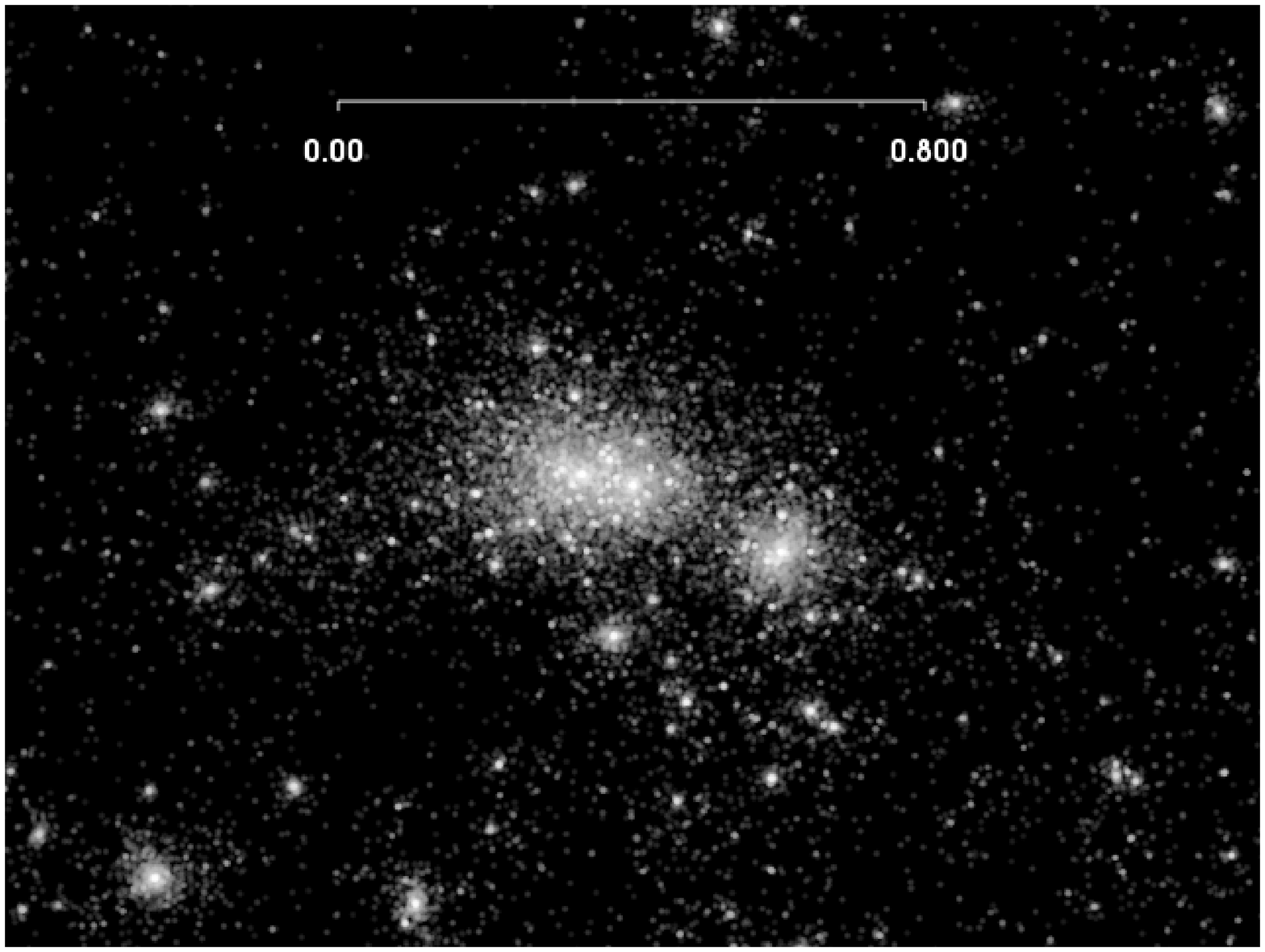}
\caption{We show images of our three Milky Ways. From top to bottom, MW.1 ($1.82 \times 10^{12} M_\odot$) from Run C, MW.2 ($0.87 \times 10^{12} M_\odot$) and MW.3 ($1.32 \times 10^{12} M_\odot$) from Run D. The left panels show both the dark (white) and luminous (red) pre-reionization halos. The right panels show only the luminous pre-reionization halos in greyscale with the brightest pre-reionization halos in white. In the right panels our Milky Ways have been rotated $\sim180^o$ relative to the view in the left hand panels. The bar across the top of each panel shows the scale in Mpc}
\label{MW1}
\end{figure*}

Run A, with a minimum particle mass of a $3.5 \times 10^6 M_\odot$, was not able to resolve subhalos within 200 kpc of the Milky Way mass halos. In Run C, we increase our mass resolution to $3.5 \times 10^5 M_\odot$ by increasing the number of pre-reionization halos in the initial conditions. By decreasing our minimum pre-reionization halo mass to $3.5 \times 10^{5} M_\odot$ we are able to resolve subhalos at $R>50$~kpc (see Figure~\ref{MW1}). At $z=0$, a luminous pre-reionization halo, evolving in isolation, is surrounded by a cloud of lower mass pre-reionization halos and tracer particles. The number of dark particles increases with the total mass of the luminous pre-reionization halo  and the mass resolution of the simulation. The detectability of the lowest mass halos by AHF is dependent on the ability of the luminous pre-reionization halos to accrete and retain their clouds of tracer particles. The larger number of low mass pre-reionization halos and tracer particles in runs C and D will allow more pre-reionization halos to accrete large enough clouds to be detected as a present day halo.

Resolving subhalos near a large galaxy is complicated by the background density field of the host halo and the stripping of the clouds of tracer particles during tidal interactions. To resolve a subhalo in the inner 100 kpc of Milky Way mass halo, the pre-reionization halo must retain enough of its cloud to be considered a bound system. In addition, it must have a high enough central density to be seen against the background of the host halo. The effect of the larger mass of the pre-reionization halo on the central concentration of the subhalo will be discussed in \S~\ref{SEC.soft}. For AHF, the lower limit to robustly detect halos at $z=0$ is a cloud of $\sim 50$ tracer particles \citep{KnollmannKnebe:09}.

Our simulations cannot provide information on the $z=0$ stellar properties of a halo for a pre-reionization halo which has undergone significant tidal disruption. Beyond the stripping of the accumulated dark cloud described above, our simulations do not allow for the breaking apart of the pre-reionization halos. We only consider a pre-reionization halo unaffected by tides if its cloud of dark particles remains intact and detectable. This negates comparisons within 50 kpc of a host halo where a significant number of the observed ultra-faint dwarfs have been modified by tides and our luminous pre-reionization halos are stripped of their clouds of tracer particles. In the galactocentric radial distributions and luminosity functions presented in Paper II of this series, we only include the simulated and observed sample at $R>50$~kpc.

\subsection{Softening Length}
\label{SEC.soft}

In this section, we discuss one of the most prevalent numerical effects of using a spectrum of particle masses in our high resolution region instead of a uniform particle masses. The effects of this spectrum of masses primarily manifests in the lower mass halos and are sensitive to our choice of the  softening length, $\epsilon$. We also show the mass functions from Runs C and D for our chosen softening length.

Typically, the softening length is set at $2\%$ of the average distance between particles in co-moving coordinates. For a representative volume of the universe with particles of uniform mass, $\epsilon = 0.02 N^{-1/3}$~Mpc, where $N$ is the number of particles per Mpc$^3$. For the high resolution region, we have a particle mass range between $3.5 \times 10^{5} M_\odot - 2.5 \times 10^{8} M_\odot$, requiring particle softening lengths from $0.5$~kpc to $2$~kpc. The pubic version of Gadget 2 does not have the capability of assigning softening lengths to each particle. Therefore, we must choose a single softening length for all the particles in the high resolution region. To determine the optimal value of $\epsilon$, we have run the same initial conditions with softening lengths in our high resolution region of $\epsilon = 0.1$~kpc, $1$~kpc and $5$~kpc. We find that the best results for $\epsilon = 1$~kpc (corresponding to a uniform particle mass of $\sim 10^7 M_\odot$).

\begin{figure*}
\plottwo{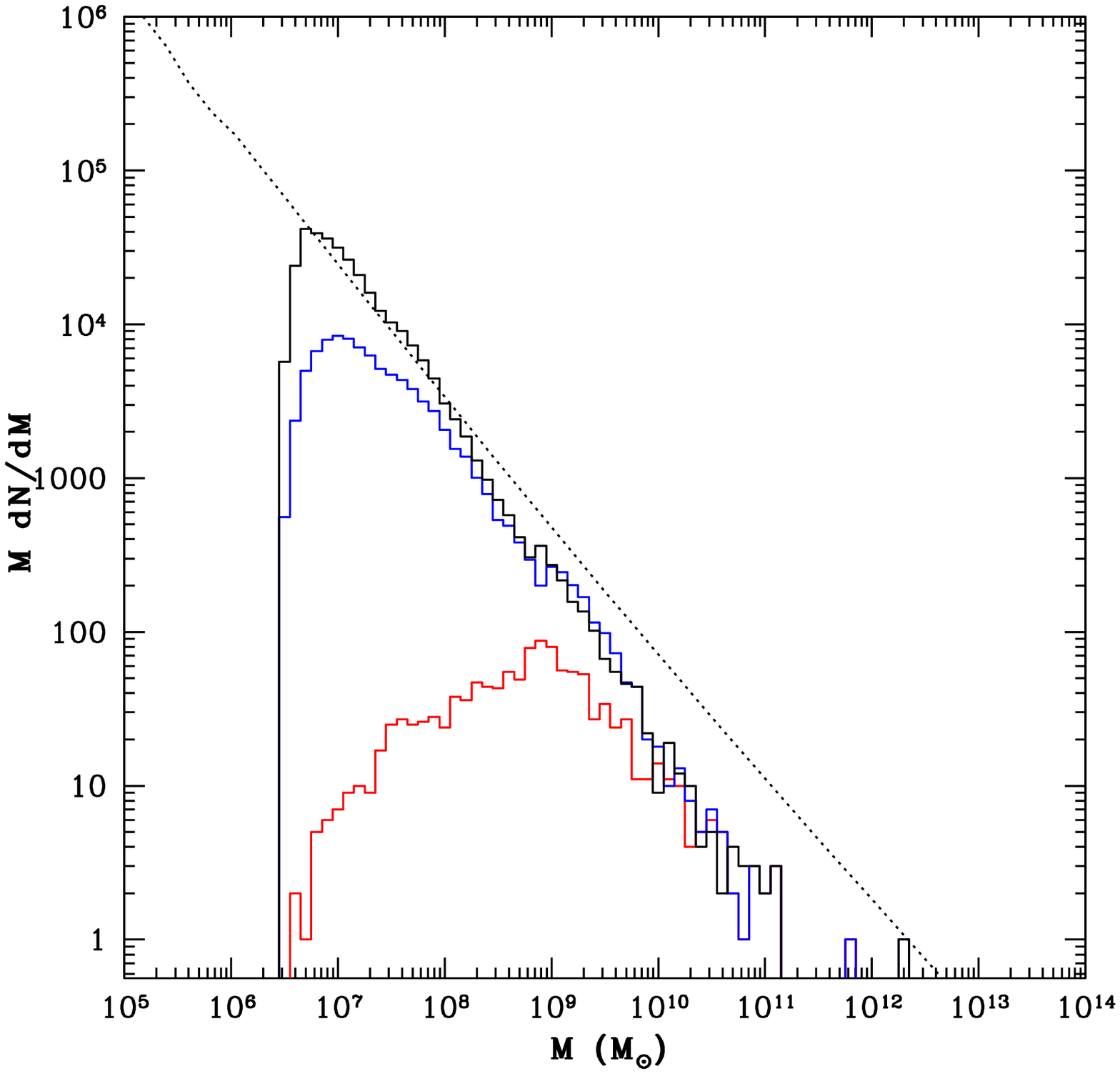}{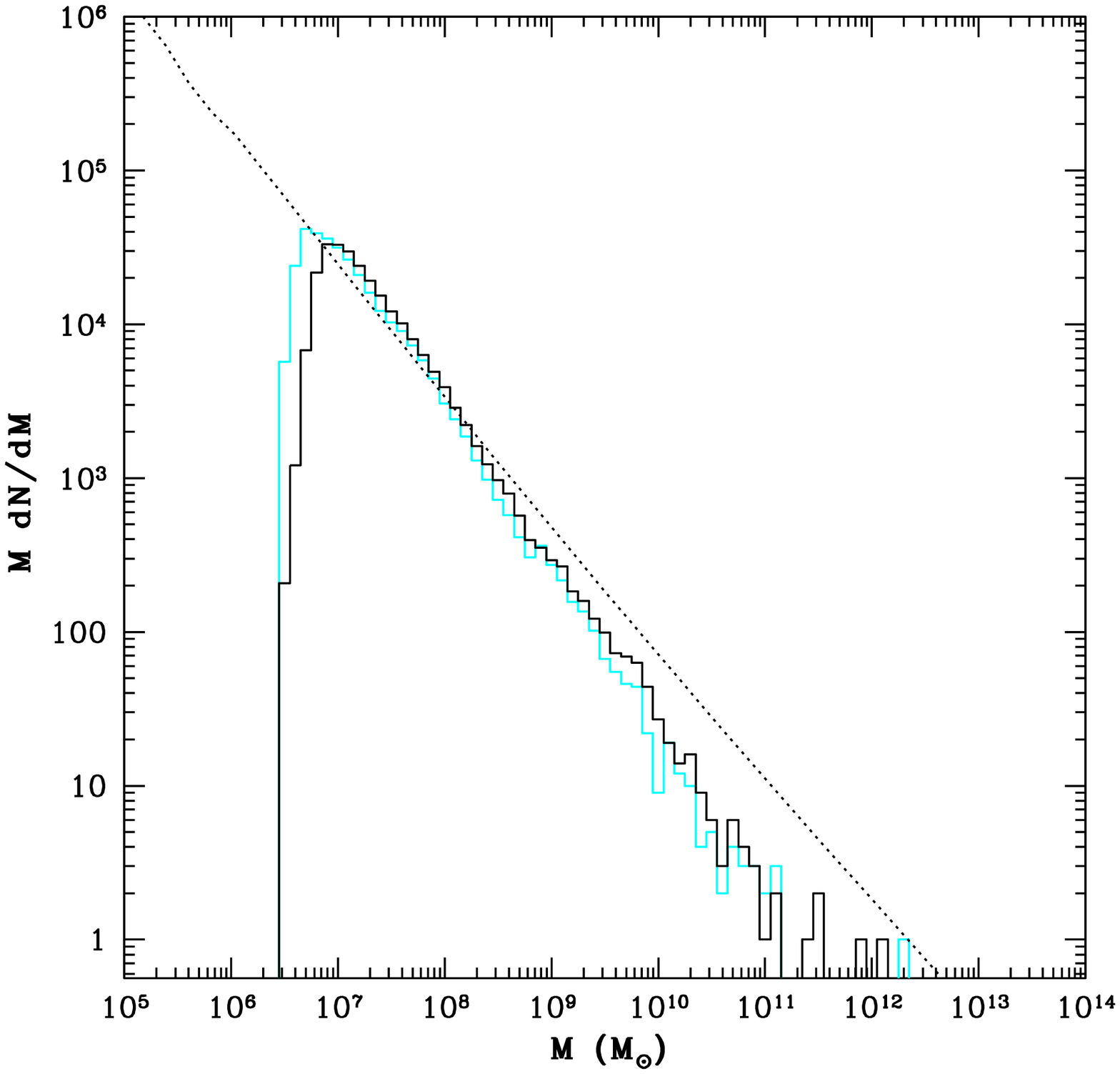}
\caption{{\it{Left}} : Mass functions for Run C evolved with three different softening lengths, 100 pc (red), 1 kpc (black) and 5 kpc (blue). Note that while there is a negligible difference at large masses, $\epsilon = 1$~kpc gives us the least residual when compared to the expected mass function. The dotted line is the Press-Schechter for a $\sim7$~Mpc$^3$ volume, equivalent to the mass of the bound halos. {\it{Right}} : Mass function of all halos found by AHF in our high resolution region for run C (gray line) and run D (black line). The dotted line is the Press-Schechter for a $\sim7$~Mpc$^3$ volume at $z=0$, equivalent to the mass of the bound halos. In both mass functions we only include $z=0$ halos which contain only high resolution particles.}
\label{MF.soft}
\end{figure*}

The right panel of Figure~\ref{MF.soft} shows the mass functions of Runs C and D compared to the  Press-Schechter mass function run with $\epsilon = 1$~kpc . For C and D we see a deficit in the number of $10^9 - 10^{11} M_\odot$ halos when compared to the Press-Schechter,and an over abundance of $M < 10^7 M_\odot$ halos. The deficit for larger halos may result from the location of our high resolution region. The Press-Schechter is the mass function of a typical volume of the universe. Our high resolution region is under-dense, containing three filaments bordering a void. The overabundance for $M < 10^7 M_\odot$ halos has a slope similar to the initial halo mass function from the pre-reionization simulations. At those masses, the $z=0$ halos are dominated by one pre-reionization halo. This suggests that the steeper slope of the mass function at low masses is a numerical effect reflecting the behavior of the $z=8.3$ mass function from the pre-reionization simulations.

When $\epsilon$ is set lower than $1$~kpc (red curve in Figure~\ref{MF.soft}), low mass halos with one or more luminous pre-reionization halo are preferentially destroyed by numerical effects. Statistically, luminous pre-reionization halos are more massive than their dark counterparts, hence they migrate to the centers of their modern halos via dynamical friction. Any two body interaction between a luminous pre-reionization halo and lower mass dark tracer particle will result in artificial heating. 
Over the entire simulation, such interactions artificially heat the cloud of tracer particles until is disperses. We find that for $\epsilon = 0.1$~kpc only the most massive pre-reionization halos with the deepest potentials are able to retain their clouds. Isolated pre-reionization halos are surrounded by an extremely tenuous cloud of low mass dark particles, which is not detected by AHF as a bound halo. 

Using $\epsilon>1$~kpc also artificially decreases the number of the low mass halos (blue curve on Figure~\ref{MF.soft}). Unlike the deep potentials of the massive halos, the potentials of halos with masses $M < 10^8 M_\odot$ are relatively shallow. If $\epsilon$ is too large, the low mass potentials will be flattened to the point where the pre-reionization halos are unable to accrete the tracer particles required for AHF detection. In halos with $M \simgt 10^9 M_\odot$, this effect is minimal. However we are primarily interested in halos with $M < 10^9 M_\odot$.

\subsection{Subhalo Scale Comparisons}
\label{SEC.subhalo}

In this section, we study the distribution of subhalos around our Milky Ways. We use runs C and D to explore the simulated distribution of $z=0$ subhalos around our Milky Way mass hosts. Comparisons are made with other CDM simulations and with observations.

In each simulation, we search for Milky Way type halos, using observational and theoretical constraints. This gives us a range of halo masses for candidate Milky Ways of $\sim0.6 \times 10^{12} M_\odot - 4 \times 10^{12}$ \citep{Watkinsetal:10,Kallivayaliletal:09,Zaritskyetal:89,Klypinetal:02}, and upper mass estimates for the Local Group of $\sim 5.3 \times 10^{12}$ \citep{LiW:08,vanderMarelG:08}. These criteria gives us three Milky Ways, one in the Run C and two in the Run D, respectively (Table~\ref{TAB.MW}). All three hosts have masses on the low end of the observed Milky Way mass range.

The Milky Way halo in Run C, MW.1, is in one of the highest density regions of our volume, with a companion galaxy of mass $10^{11} M_\odot$ at a distance of 2 Mpc. The Milky Ways in Run D have masses of $0.87 \times 10^{12} M_\odot$ for MW.2 and $1.32 \times 10^{12} M_\odot$ for MW.3. Though they are both in filaments, the nearby environments of MW.2 and MW.3 differ (see Figure~\ref{LSS}). MW.2 sits at the intersection of three filaments, and there are $\sim 10^{11} M_\odot$ halos within 1.5 Mpc. In contrast, MW.3 is only 1-2 Mpc away from a complex of galaxies with masses $\sim 10^{11} M_\odot$ that appears to be in the process of merging to form another Milky Way mass system. For our comparisons with traditional simulations, and with observations, we use all three Milky Way mass halos. This allows us to explore differences between the first and second order as well as variations introduced by environmental effects.

\begin{table}
\centering
\begin{tabular}{ c c c c c }
\hline
Name & Run & Mass & $R_{vir}$ & $v_{max}$ \\
& & $(10^{12} M_\odot$) & (kpc) & (km~s$^{-1}$) \\
\hline
\hline
MW.1 & C & 1.82 & 248.1 & 203.4 \\
MW.2 & D & 0.87 & 222.6 & 196.6 \\ 
MW.3 & D & 1.32 & 194.7 & 177 \\ 
\hline
\end{tabular}
\caption{Table of the Milky Way mass halos. The columns from left to right are (1) Milky Way identifier, (2) simulation run each Milky Way is embedded in, (3) the AHF derived virial mass of each halo in $10^{12} M_\odot$, and (4) the AHF derived $v_{max}$ of the halo in km~s$^{-1}$.}
\label{TAB.MW}
\end{table}

\begin{figure}
\plotone{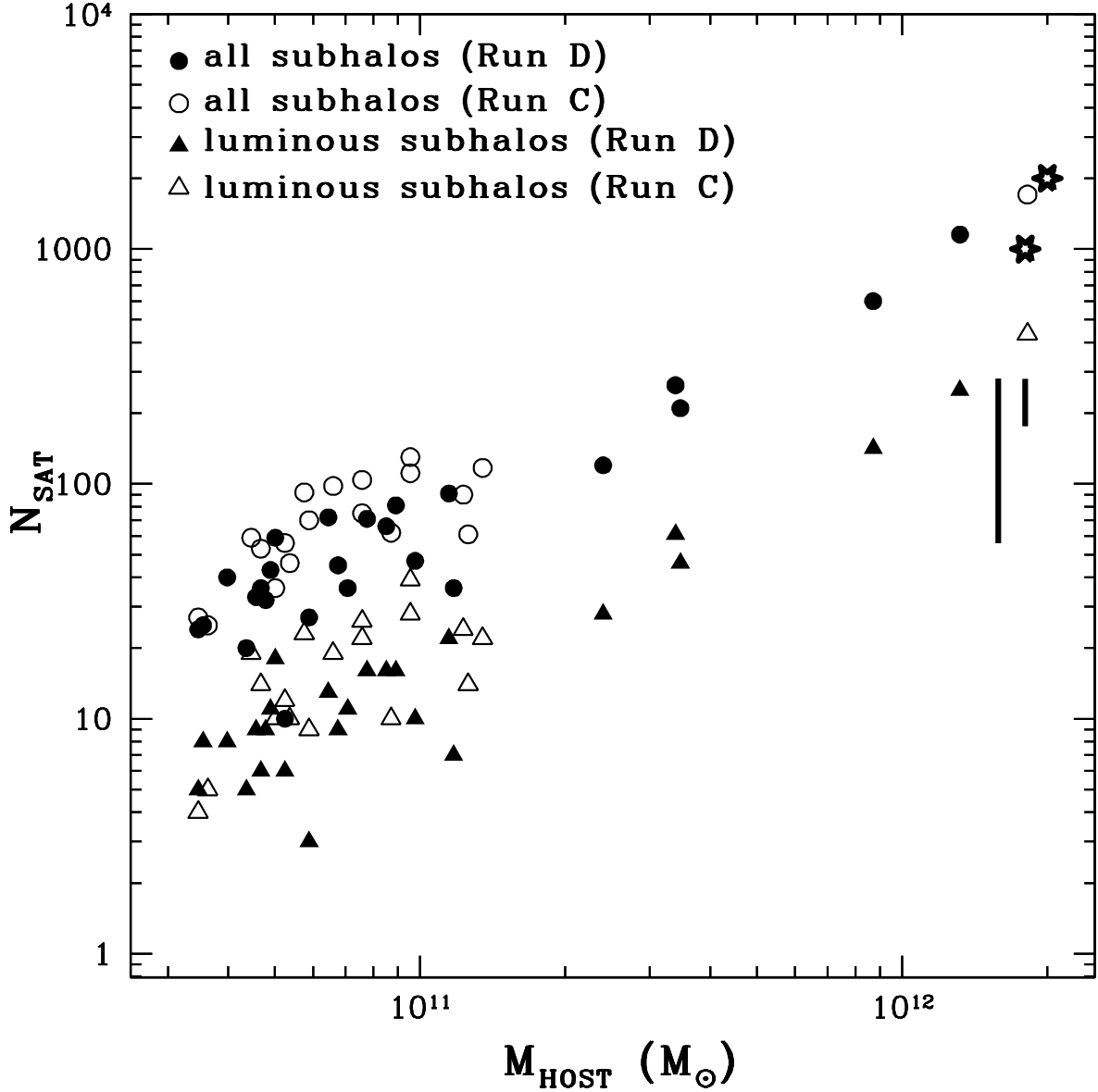}
\caption{Number of satellites as a function of the mass of the host halo.  The total number of satellites within $R_{vir}$ for each halo are represented by circles, and the number of luminous satellites within $R_{vir}$ for each halo by triangles. The results from runs C and D are shown as the opened and filled symbols respectively. The predictions from Via Lactea II \citep{Diemandetal:08} and Aquarius \citep{Springeletal:08}, scaled to our mass resolution, are shown as the opened stars. Aquarius is the star with a greater number of satellites within $R_{vir}$. The ranges of the \cite{Tollerudetal:08} and \cite{Walshetal:09} predictions at 200 kpc are the green and purple barred lines, respectively.}
\label{NSAT}
\end{figure}

Before looking at the distribution of satellites around individual Milky Ways, we check the distribution of the number of dark matter subhalos as a function of host mass. In Figure~\ref{NSAT}, we show a linear relation between host mass and the number of satellites for both Runs C and D. There is good agreement with the Via Lactea and Aquarius runs when we adjust their results for our lower mass resolution. Our simulations can robustly resolve halos with $M > 10^7 M_\odot$ ($v_{max} \simgt 5.5$~km~s$^{-1}$). To scale the number of subhalos within $R_{vir}$ in the Via Lactea and Aquarius simulations, we use  $v_{max}\sim5$~km~s$^{-1}$ for Via Lactea and $v_{max}\sim7$~km~s$^{-1}$ for Aquarius (from Figure 27 in \cite{Springeletal:08}). 

We use knowledge of the stellar properties of the pre-reionization halos to investigate the expected number of luminous satellites for a given host mass. We consider a subhalo luminous if it contains at least one pre-reionization halo with $M_{\ast}>10^2 M_\odot$, or has a $z=0$ mass $M>10^9 M_\odot$ ($v_{max}>20$~km~s$^{-1}$). To study the distribution of the number of luminous satellites, $N_{sat} (L_V > 10^2 L_\odot)$ vs. $M_{host}$ we do not need to know the luminosity of the satellite at $z=0$, only whether it is luminous. We find all of the luminous subhalos in Figure~\ref{NSAT} formed stars before reionization since we have no $z=0$ halos above the threshold for post-reionization gas accretion ($10^9 M_\odot  : v_{max} = 20$~km~s$^{-1}$) which do not contain a primordial stellar population. For Runs C and D, we find the number of luminous satellites increases linearly with host mass. For hosts with $M < 10^{11} M_\odot$, we see a larger scatter in the total number of satellites. Additionally, in that host mass range, we see greater scatter in the mapping of the total number of satellites to the number of luminous satellites. Since Runs C and D contain only three Milky Way mass systems, the lack of scatter may also be due to small number statistics. The decrease in scatter may be a function of how dominant the halo is in its environment. In the filaments, a $10^{12} M_\odot$ halo dominates the region around it, negating any environmental effects inside the virial radius. A lower mass host, however, is not able to dominate its environment. Therefore, the number of satellites for low mass hosts will be more sensitive to the environment in which they are embedded.

The current observational sample of dwarfs is complete only to within 50 kpc \citep{SimonGeha:07,Koposovetal:07,Walshetal:09}. The corrections for the detection limits of current surveys were done from a theoretical perspective by \cite{Tollerudetal:08}. They used halos from Via Lactea I \citep{Diemandetal:07a}, assuming a simple relationship between halo mass and luminosity for the subhalos. The range of \cite{Tollerudetal:08} is shown on Figure~\ref{NSAT} as the shorter, thick black line. Unlike their work, our simulations do not assume a relationship of luminosity to halo mass. Instead, we draw the stellar properties of the $z=0$ halos directly from the cosmologically consistent pre-reionization simulations. This accounts for the large scatter in stellar mass as a function of halo mass for the smallest galaxies \citep{RicottiGnedinShull:02b}. Our results are consistent with the upper end of the \cite{Tollerudetal:08} range for the number of luminous satellites within $\sim 200$~kpc. Based on these comparisons, the total number of subhalos and number of luminous satellites around MW.1, MW.2, and MW.3 are in agreement with results of other published works. 

\begin{figure}
\plotone{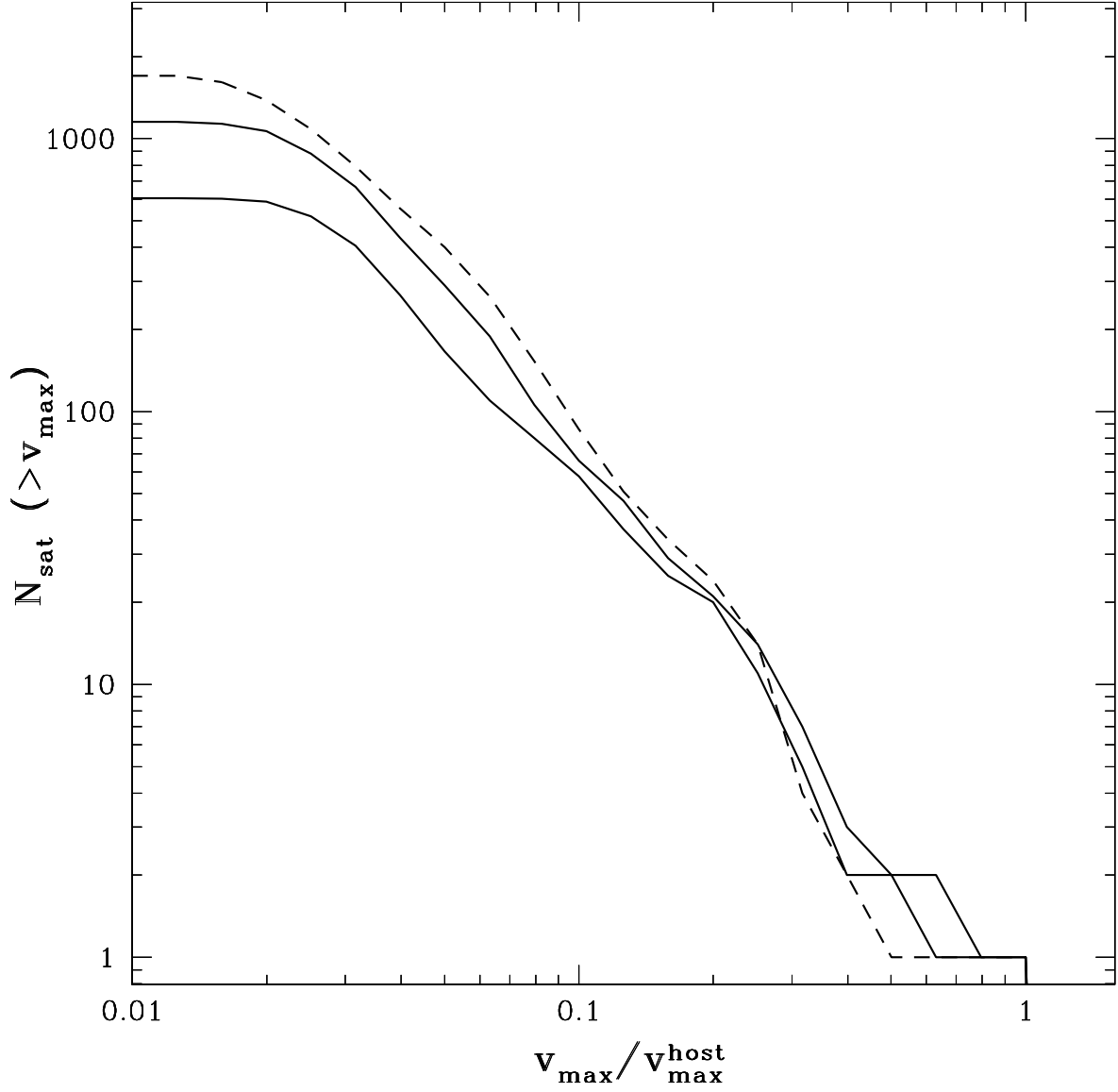}
\caption{Number of satellites within $R_{vir}$ with greater than a given $v_{max}$ for our two second-order (black lines) and one first-order (gray line) Milky Ways. The two versions of our method produce equivalent distributions and match the CDM simulations from \cite{PolisenskyRicotti:10}.} 
\label{VCIRC}
\end{figure}

We next compare the distribution of maximum circular velocity for all subhalos around a Milky Way for our simulations with other CDM simulations. We find that the satellite mass functions for halos from Runs C and D are consistent with one another, and results from Aquarius, Via Lactea and \cite{PolisenskyRicotti:10} (Figure~\ref{VCIRC}). Based on this, we argue that our simulations can reproduce the number and distribution of subhalos around the Milky Ways, as well as traditional N-body simulations. In the next section, we discuss the observational and theoretical criteria for a halo to be defined as a fossil of the first galaxies.

\section{Results}
\label{SEC.Res}

\subsection{Definition of a Fossil Dwarf}

For observed dwarfs, a fossil is defined as a dSph which underwent $\simgt 70\%$ of its star formation before reionization and today is a diffuse, spherical system devoid of gas \citep{RicottiGnedin:05}. These dim dwarfs populate dark matter halos whose circular velocities have never been above the filtering velocity, preventing them from accreting gas from the IGM after reionization. In our simulations,we define a fossil halo for which $max(v_{max}(z))<v_{filt}$. Any halo with $v_{max}(z = 0) < v_{filt}$ is referred to as a {\it{candidate fossil}}. However, in regimes where tidal stripping is considerable, there is a significant chance that a halo with a $v_{max}<v_{filt}$ at $z=0$ had a maximum circular velocity above the threshold for accretion from the IGM at an earlier time \citep{Kravtsovetal:04}. 

Given these criteria, we classify our $z=0$ halos into three populations as follows. (1) A {\it{non-fossil}} is a $z=0$ halo for which $v_{max}(z = 0) > v_{filt}$. (2) Halos which are candidate fossils but for which $max(v_{max}(z))$ was above the IGM accretion threshold in the past are classified as {\it{polluted fossils}}. The non-fossils and a fraction of the polluted fossils accreted gas from the IGM and formed a significant population of stars after reionization. Therefore, our simulations cannot provide robust information on the non-fossil and polluted fossil stellar properties in the modern epoch. (3) For the {\it{true fossils}} we are able to generate detailed information on their stellar properties. A true fossil is defined as any $z=0$ halo for which $v_{max}$ never exceeded the IGM filtering mass, suppressing gas accretion and star formation after reionization. 

\begin{figure}
\plotone{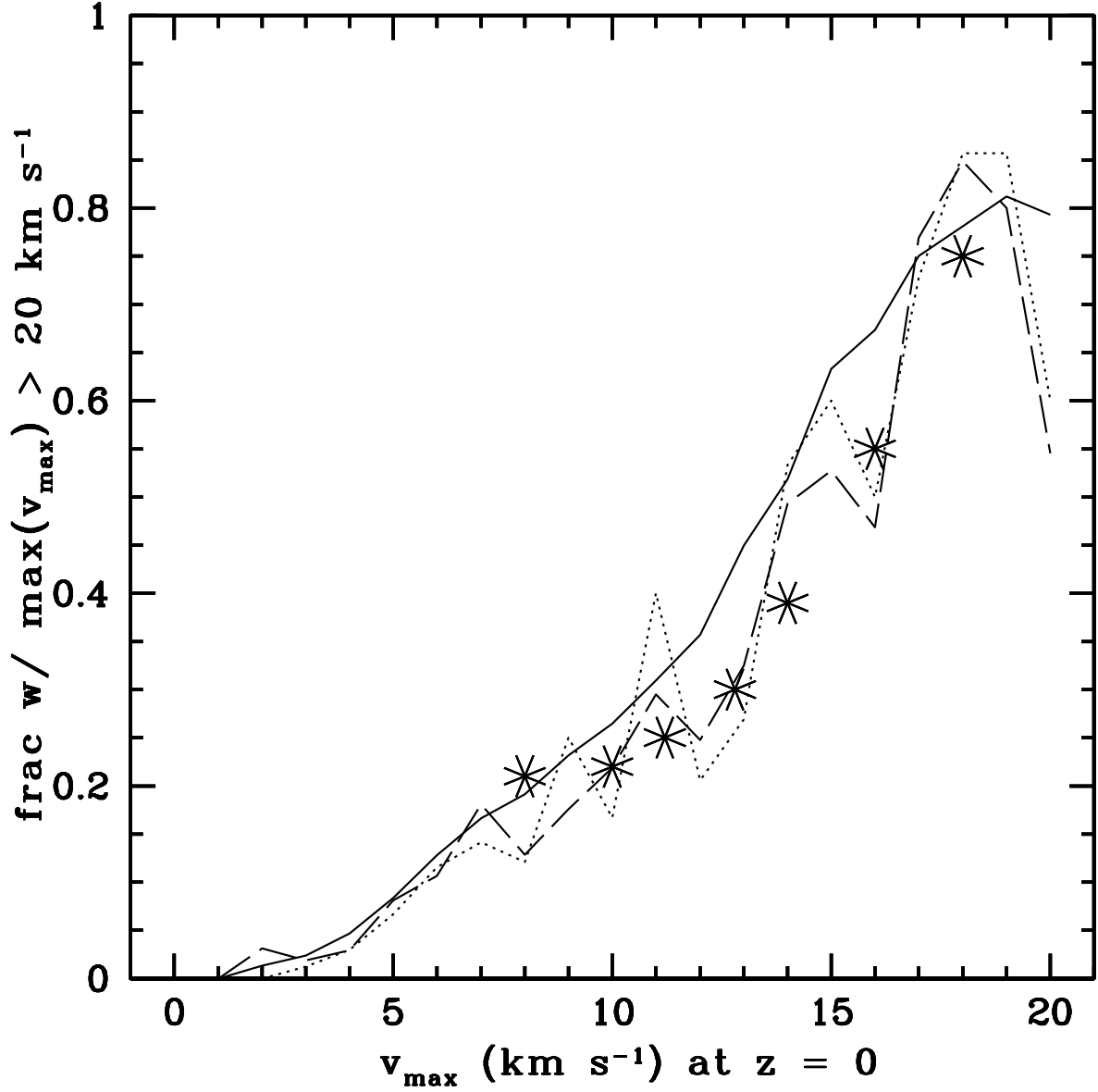}
\caption{Fraction of candidate fossils with max$(v_{max}(z))>v_{filt}$ where $v_{filt} = 20$~km~s$^{-1}$ for Run D (lines) and \cite{Kravtsovetal:04} (asterisks). The solid, dashed and dotted lines show the fraction of true fossils for three different sub-populations. The solid line shows the relation for all the candidate fossils in Run D, while the dashed and dotted lines show the fraction of true fossils for candidate fossils within $1$~Mpc and $400$~kpc of MW.3 respectively.}
\label{VMAX.frac}
\end{figure}

To separate the polluted fossils from the true fossils of the first galaxies, we follow the $v_{max}$ evolution for each candidate fossil back from $z=0$ to $z_{init}$. We find that $f(v_{max})$, the fraction of candidate fossils which have $max(v_{max}) > v_{filt}$, as a function of their $v_{max}(z=0)$, is consistent with results found by \cite{Kravtsovetal:04} (see Figure~\ref{VMAX.frac}). In addition, we find that $f(v_{max})$ does not have a strong dependence on the environment of the fossils.  When we compare the results for all the fossils (solid line) with those within 1 Mpc (dotted line) and 400 kpc (dashed line) of MW.2 and MW.3 we do not see a significant difference. These results are independent of the choice of the filtering velocity. For the remainder of this work, we use the term fossil in reference to only these true fossils.

In addition to maintaining $v_{max} < v_{filt}$ for its entire evolution, a fossil must also survive to $z=0$ without being tidally stripped. While objects which have undergone tidal stripping are unlikely to retain their pre-reionization stellar properties \citep{PenarrubiaNM:08}, our initial conditions do not allow us to simulate tidal effects beyond the stripping of a $z=0$ halo's tracer particles. The use of N-body particles to represent pre-reionization halos forces the masses of those halos to be conserved. No matter how strong the tidal forces are, the stellar and dark matter properties will not change.  This is in no way consistent with the current understanding of the effect of tidal stripping on a satellite's stellar population. 

While the dark matter halo can be stripped away, leaving the stellar properties relatively intact \citep{Choietal:09,PenarrubiaMN:08}, once $90\%$ of the dark matter has been stripped and the mass loss reaches the outer stellar radii, the stripping of the stellar populations will occur at a faster rate than the denser dark matter cusp increasing the mass-to-light ratio of the system \citep{PenarrubiaNM:08}. We have no way of tracking the mass loss of an isolated pre-reionization halo to determine which components have been disrupted. We therefore err on the side of caution: we use the destruction of a $z=0$ halo's dark particle cloud to flag halos which have undergone tidal stripping. Any present day halo whose cloud of tracer particles has been destroyed or stripped down to $N \simlt 50$ particles, will not be robustly detected as substructure and its mass will be added to that of the host galaxy. If $N < 20$, the $z=0$ halo will not be detected at all \citep{KnollmannKnebe:09}. This adds a second, lower mass criteria for a pre-reionization halo to be identified as part of a fossil. It must be ``found'' in a halo at $z = 0$ by AHF to be considered a fossil. Any pre-reionization halo not in a $z=0$ halo is assumed to be completely disrupted.

\begin{figure}
\plotone{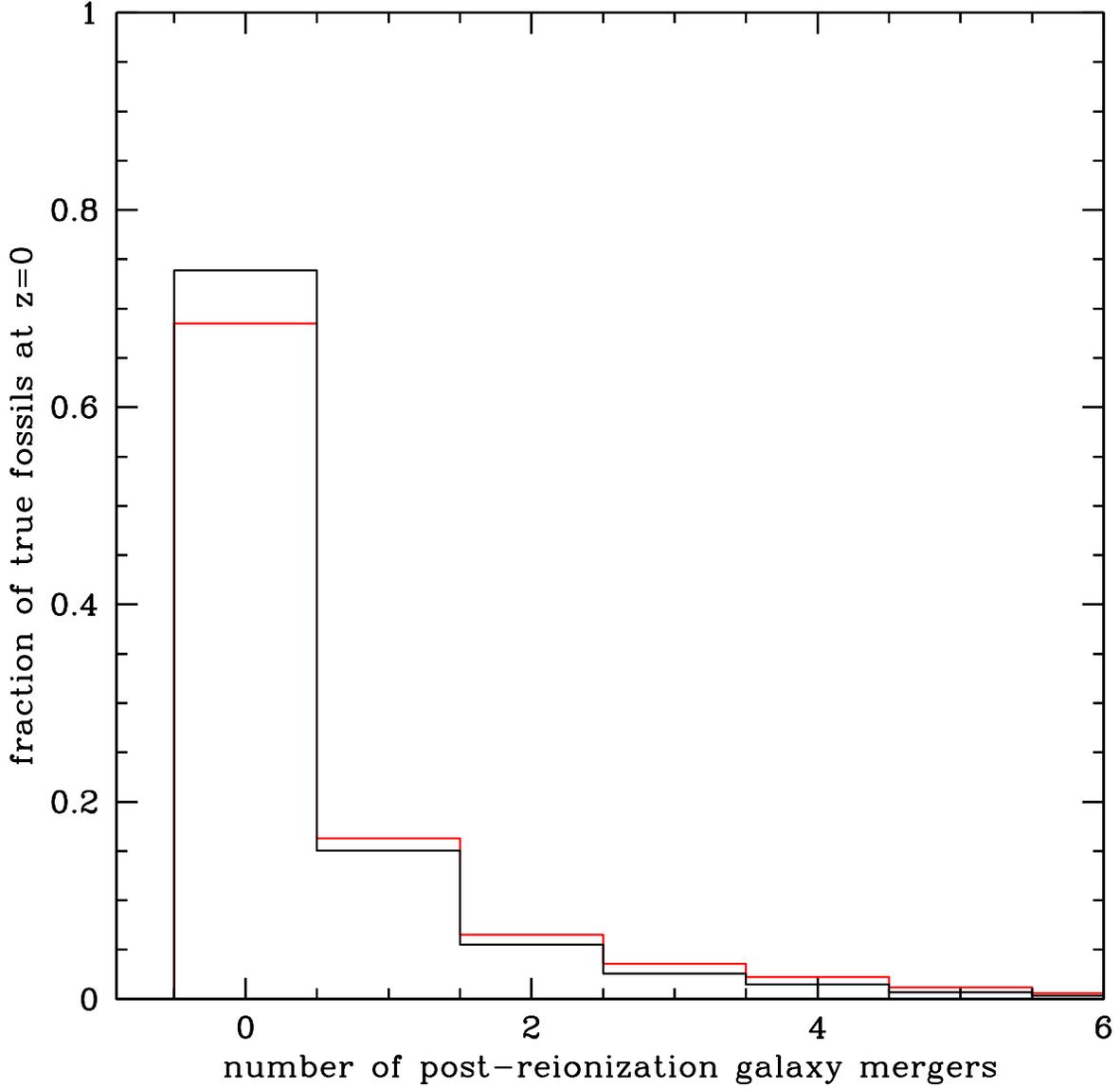}
\caption{Fraction of luminous true fossils which have undergone $<6$ galaxy mergers after reionization for $v_{filt}=20$~km~s$^{-1}$ (black) and $v_{filt}=30$~km~s$^{-1}$ (red). We define a galaxy merger as any merger in which two or more of the components contain a luminous population. For $>4-5$ galaxy mergers, the fraction of $z = 0$ true fossils becomes negligible.}
\label{NLUM.fos}
\end{figure}

Given these criteria, we can say a few things about our fossil population. Our fossils are dimmer and less massive than the polluted fossils and non-fossils. As a population, they are less likely to have undergone mergers involving two or more luminous pre-reionization halos (Figure~\ref{NLUM.fos}). We define a merge between two or more luminous pre-reionization halos as a galaxy merger. Using an $v_{filt} = 20$~km~s$^{-1}$, $25\%$ of the fossils have two or more luminous pre-reionization halos compared to $40\%$ of candidate fossils. The majority of true fossils $(75\%)$ contain only one luminous pre-reionization halo, however the remainder do not represent a negligible fraction. We find the same result when using the $v_{filt}=30$~km~s$^{-1}$ adopted by GK06. As with the $20$~km~s$^{-1}$ case, $75\%$ of true fossils have only one luminous pre-reionization halo. Therefore, while the majority of fossils have not undergone galaxy mergers, it is not an effect that can be ruled out.

\subsection{Luminosity Threshold for Fossils}
\label{SEC.gk06}

\begin{figure}
\plotone{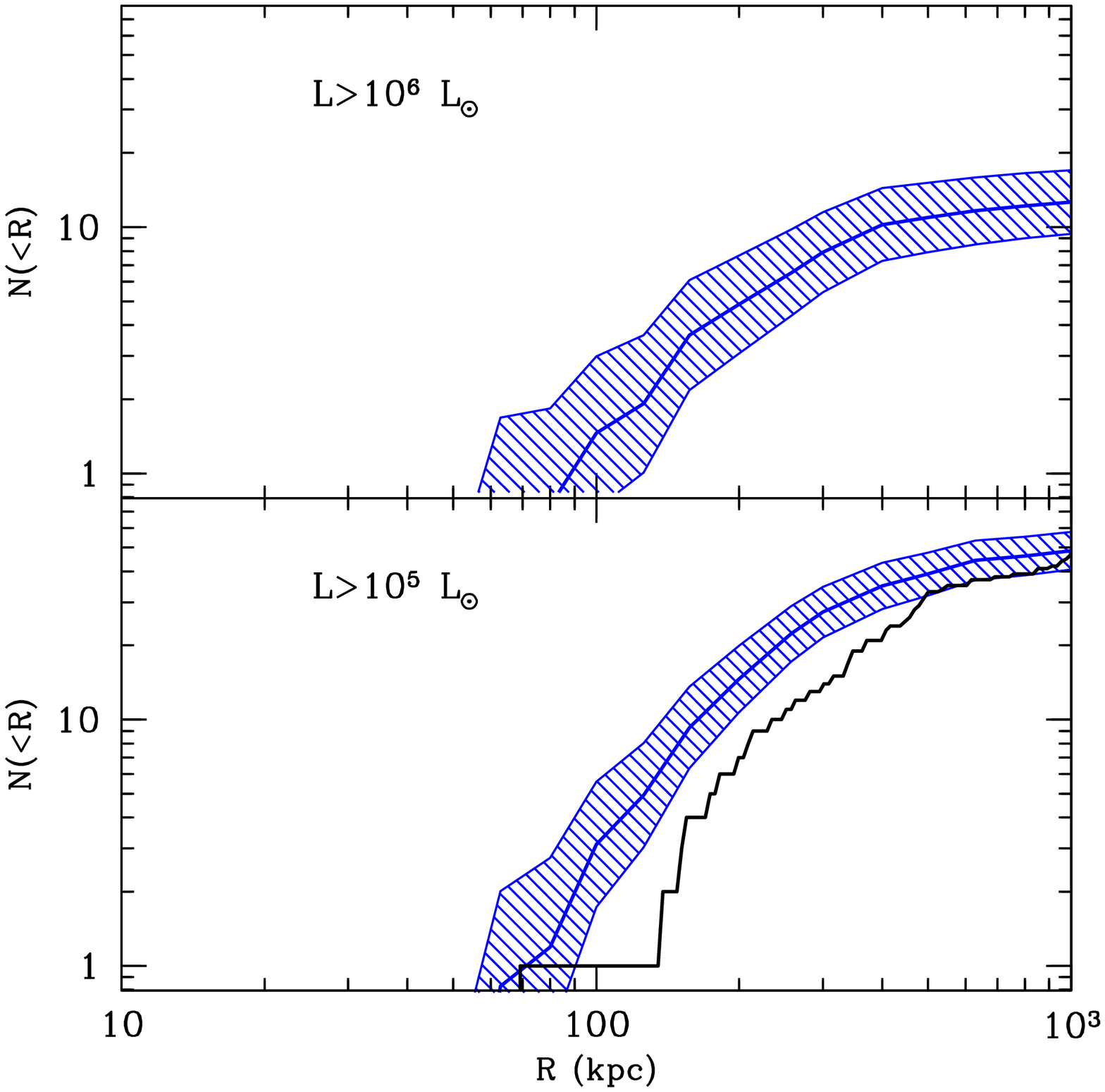}
\caption{The radial distribution of the true fossils around MW.1 in Run C (black lines) and the results from GK06 (blue band) for halos with $L_V > 10^5 L_\odot$ and $L_V > 10^6 L_\odot$. We have used a $v_{filt}=30$~km~s$^{-1}$ threshold to determine whether a $z=0$ halo is a true fossil.}
\label{RD.gk06}
\end{figure}

Before making detailed comparisons between our simulations and observations, we compare our work and the N-body simulations in GK06. Unlike our method, which allows us to directly trace the pre-reionization halos to the present day, GK06 statistically matches pre-reionizaion halos to their counterparts at $z=0$ based on their $v_{max}$ at $z=8.3$. To make a direct comparison with GK06 we must use our MW.1 from Run C, since GK06 only used the $z=8.3$ outputs from the pre-reionization simulations. 

Figure~\ref{RD.gk06} shows the galactocentric radial distribution for GK06 (blue band) and for MW.1 (black lines). Both curves only include the true fossils. For $L_V>10^5 L_\odot$ (lower panel) we find that our simulations are consistent with GK06, if on the low end of their range. However, the brightest simulated true fossils in GK06 with $L_V>10^6 L_\odot$ have no counterparts around MW.1. We ascribe this discrepancy to the difference in how our work follows the pre-reionization halos to the modern epoch. 

\begin{figure*}
\plottwo{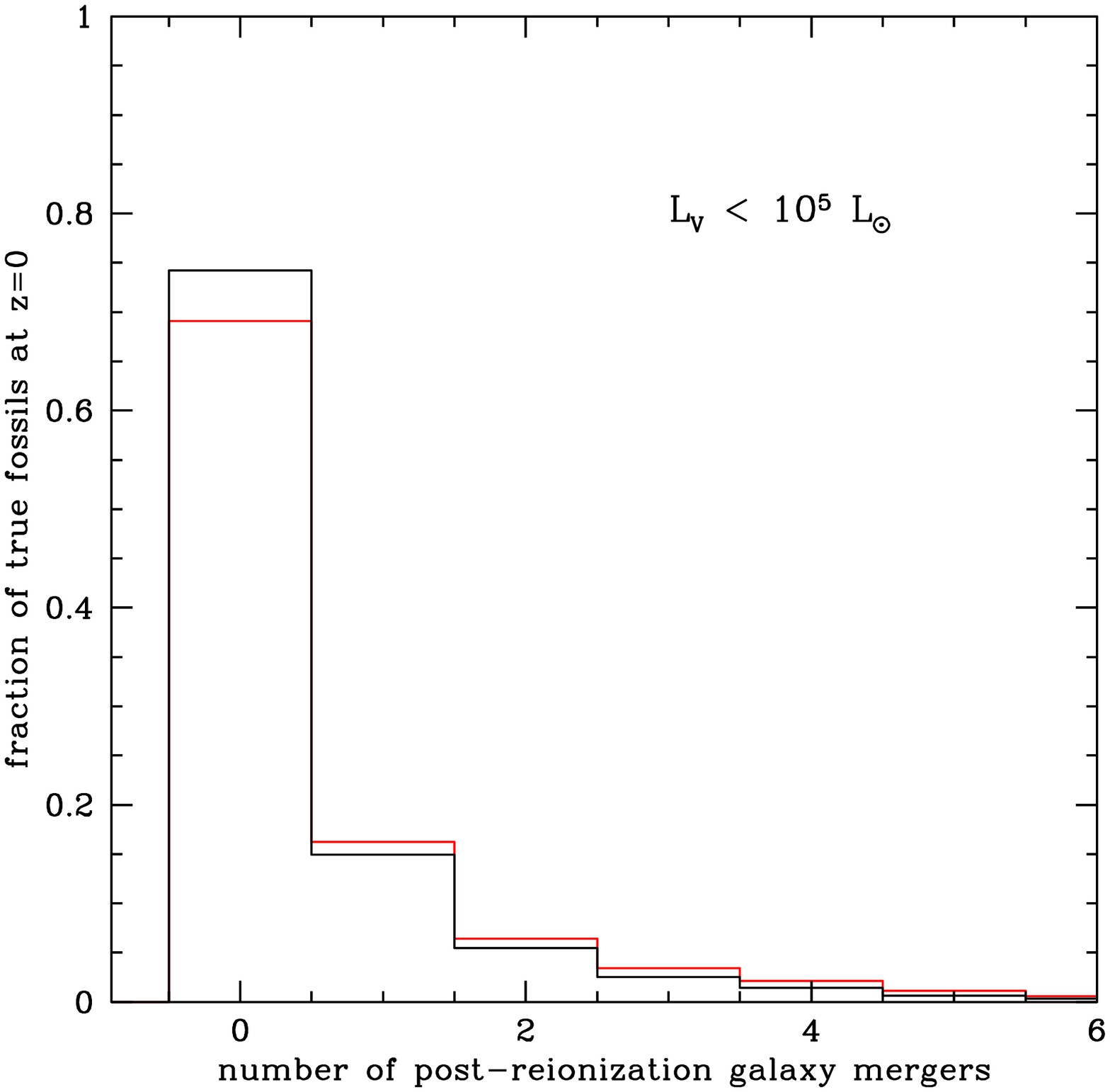}{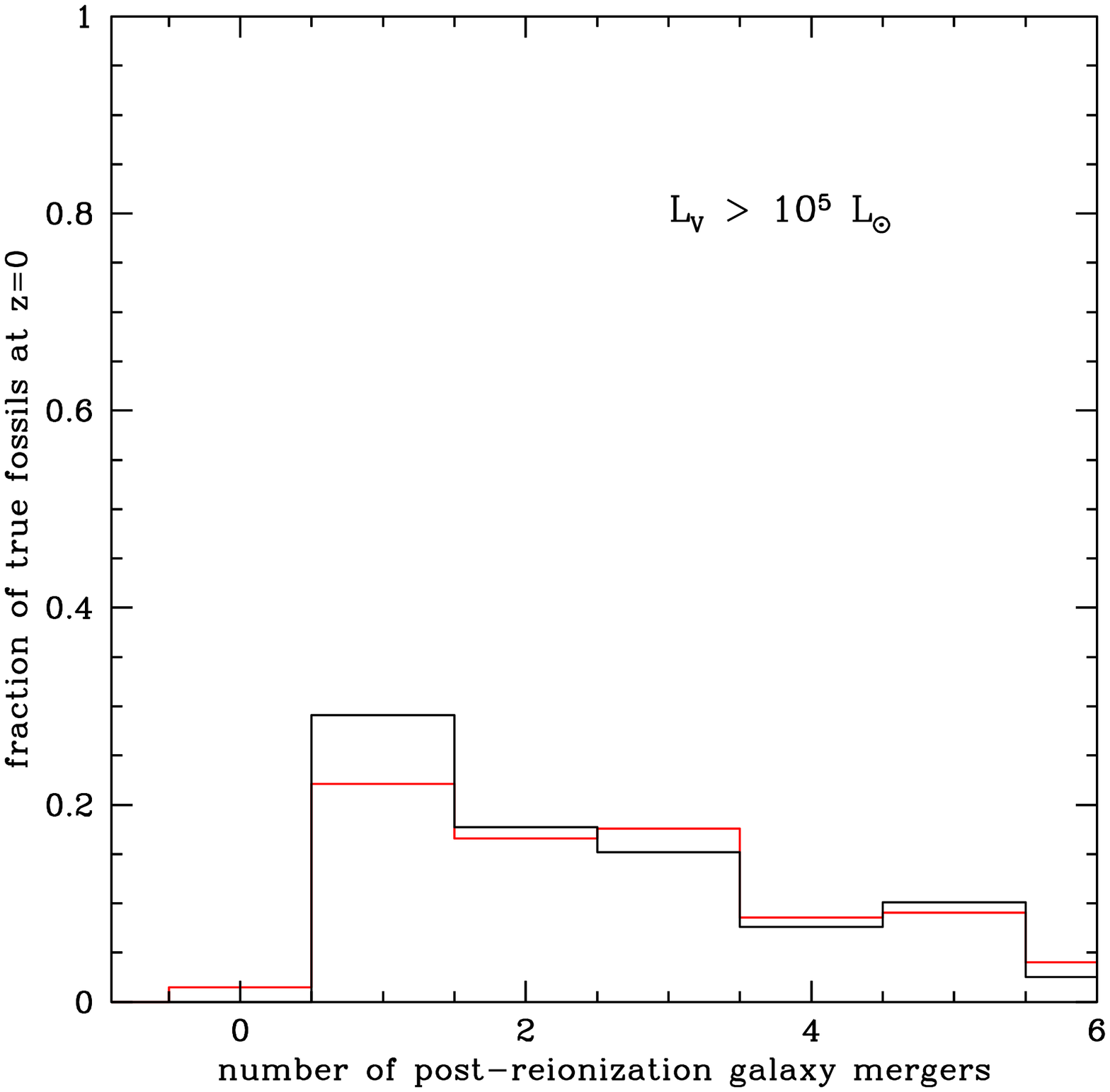}
\caption{{\it{(Left)}} Histogram of the fraction of true fossils at $z=0$ with a given number of galaxy mergers after reionization for $v_{filter}=20$~km~s$^{-1}$ (black line) and $v_{filter}=30$~km~s$^{-1}$ (red line). As in Figure~\ref{NLUM.fos}, the number of galaxy mergers is a proxy for the number of luminous pre-reionization halos in a $z=0$ halo.  In this panel we show only the true fossils with $L_V(z=0)<10^5 L_\odot$. {\it{(Right)}} The fraction of true fossils with a given number of luminous pre-reionization halos for only those with $L_V(z=0) > 10^5 L_\odot$ for $v_{filter}=20$~km~s$^{-1}$ (black line) and $v_{filter}=30$~km~s$^{-1}$ (red line). Note the shifted peak and different shape of the histogram in this panel.}
\label{NLUM.lv}
\end{figure*}

While both methods allow for the growth and stripping of a halo via accretion and tidal forces, our simulations also account for clustering of the pre-reionziation halos. The most luminous pre-reionization halos correspond to the most massive halos at $z=8.3$. These $10^7-10^8 M_\odot$ galaxies are preferentially located in higher density regions within the $1$~Mpc$^3$ pre-reionization simulation. This increases the probability that the pre-reionization halos with $L_V>10^5 L_\odot$ will have undergone a galaxy merger relative to those with $L_V<10^5 L_\odot$. In Figure~\ref{NLUM.lv}, we show the histogram of the number of luminous pre-reionziation halos for true fossils with $L_V<10^5 L_\odot$ (left panel) and $L_V>10^5 L_\odot$ (right panel). Only $\sim0-5\%$ of the highest luminosity fossils have never undergone a galaxy merger compared to $\sim90\%$ of fossils with $L_V<10^5 L_\odot$. This is independent of our choice of filtering velocity.

Why does this explain the discrepancy between our results and GK06 in Figure~\ref{RD.gk06}? The definition of a true fossil is a dwarf whose maximum circular velocity has never gone above the threshold for accretion for the IGM. In Figures~\ref{RD.gk06}~and~\ref{NLUM.lv}, we set $v_{filt}=30$~km~s$^{-1}$. Since the brightest pre-reionization halos are also the most massive, one or two galaxy mergers at high redshift would be enough to push $v_{max}$ above the filtering velocity and classify the halo as a non-fossil. In Run C, there are only 11 true fossils with $L_V>10^6 L_\odot$, none of which are within $1$~Mpc of MW.1.

{\it{This gives us a maximum luminosity threshold, $10^6 L_\odot$, above which an observed dwarf is unlikely to be a primordial fossil.}} Of the true fossil candidates identified in RG05, this puts seven into question; And I ($4.49\times10^6 L_\odot$), And II ($9.38 \times10^6 L_\odot$), And III ($1.13\times10^6 L_\odot$), And VI ($2.73\times10^6 L_\odot$), Antila ($2.4\times10^6 L_\odot$) and KKR 25 ($1.2\times10^6 L_\odot$) around M31, and Sculptor ($2.15\times10^6 L_\odot$) around the Milky Way. The remaining seven, And V, Cetus, Draco, Phoenix, Sextans, Tucana and Ursa Minor all have $L_V < 10^6 L_\odot$ and remain reasonable candidates for the fossils of the first galaxies. The classical Milky Way fossils, Sextans and Ursa Minor, as well as the ultra-faint Canes Venatici I, all have metallicity distributions suggesting star formation durations $<1$~Gyr and populations $> 10 $~Gyr old \citep{Kirbyetal:11b}. These dwarfs, in addition to Sculptor, have star formation histories that are dominated by outflows, in contrast to their brighter counterparts (``polluted fossils'' Fornax and Leo I $\&$ II) \citep{Kirbyetal:11a}. However, unlike the other outflow dominated dwarfs which have relatively short star formation bursts, Sculptor has undergone star formation over several Gyrs \citep{Babusiauxetal:05,Shetroneetal:03,Tolstoyetal:03} and a fraction of Draco's stars may be of intermediate age \citep{CioniHabing:05}.

\subsection{Fossil Properties}
\label{SEC.prop}

In this section, we present the stellar properties of our simulated true fossils and compare them with observed stellar properties of Milky Way satellites. These comparisons include V-band luminosity, $L_V$, half-light radius, $r_{hl}$, metallicity, $[Fe/H]$, and mass inside the half-light radius $M_{1/2}$ \citep{Wolfetal:10}. In BR09, we showed strong statistical agreements between the stellar properties of the pre-reionization halos, and the observed distribution of known classical dSph and ultra-faint dwarfs. Here we improve our previous results by relaxing some of the assumptions made in BR09. 

As in GK06, BR09 assumed that none of the luminous pre-reionization halos had undergone a galaxy merger. Thus, the present day distribution of stellar properties for the fossils would be identical to that of the pre-reionization halos. In addition, our previous work assumed the voids were reheated to $T\sim10^4$~K well after the clusters and filaments, as expected for UV reionization by stars. As seen in Figure~\ref{LSS}, a universe reionized first in the clusters and filaments and then in the voids, Run C, produces a larger number of luminous objects in the voids when compared to a universe reionized at the same time by redshifted X-rays from primordial black holes \citep{RicottiOstriker:04,RicottiOstrikerGnedin:05} (Run D). As in BR09, for all observed stellar properties, we use the measurements with the lowest error bars.

From hierarchical formation models, we know that all halos have undergone merger/accretion events since their epochs of formation. For $60\%$ of our pre-reionization halos, these mergers are with dark halos, producing a daughter halo with the same stellar properties as the parent. However, for all runs and all halos, regardless of their fossil status, $\sim 40\%$ of the $z=0$ halos contain more than one luminous pre-reionization halo. These galaxy mergers will change the stellar properties of the systems. 

True fossil halos in the modern epoch derive their stellar properties solely from their pre-reionization populations. For the $75\%$ of luminous true fossils which contain only one luminous pre-reionization halo, the $z=0$ stellar properties are taken directly from those of the pre-reionization halo. We account for the reddening of the stellar population by using a $M^{rei}_\ast/L \sim 5$. Note that we use such a large stellar mass to light ratio to account for stellar mass lost since reionization. The stellar mass to light ratio of our simulated galaxies at $z=0$ is,
\begin{equation}
\frac{M_\ast^{rei}}{L} = \lgroup\frac{M_\ast^{today}}{L}\rgroup\lgroup\frac{M_\ast^{rei}}{M_\ast^{today}}\rgroup
\end{equation}
where $M_\ast^{rei}$ and is the mass of the stellar population at reionization, and $M_\ast^{today}$ is the mass of the stellar population at $z=0$. The ratio between them, $M_\ast^{rei}/M_\ast^{today}$ is between 2 and 20 depending on the IMF of primordial stellar population. RG05 used a range of $M/L$ ratios and found no dependence of the fossil properties on the choice of mass to light ratio.

For the one-quarter of true fossils which have undergone a galaxy merger, the stellar properties are calculated as follows.  Throughout this section, the superscript $f$ will denote the stellar and dark matter properties of the $z=0$ halo, and the superscript $i$ the properties of the component, luminous pre-reionization halos. 

The final V-band luminosity, $L_V^f$ of a fossil halo at $z=0$, is the sum of the V-band luminosities, $L_V^i$, of the component pre-reionization halos. We assume stellar mass is conserved during all the mergers, an assumption that will be addressed in future, higher resolution simulations.  

\begin{figure*}
\plottwo{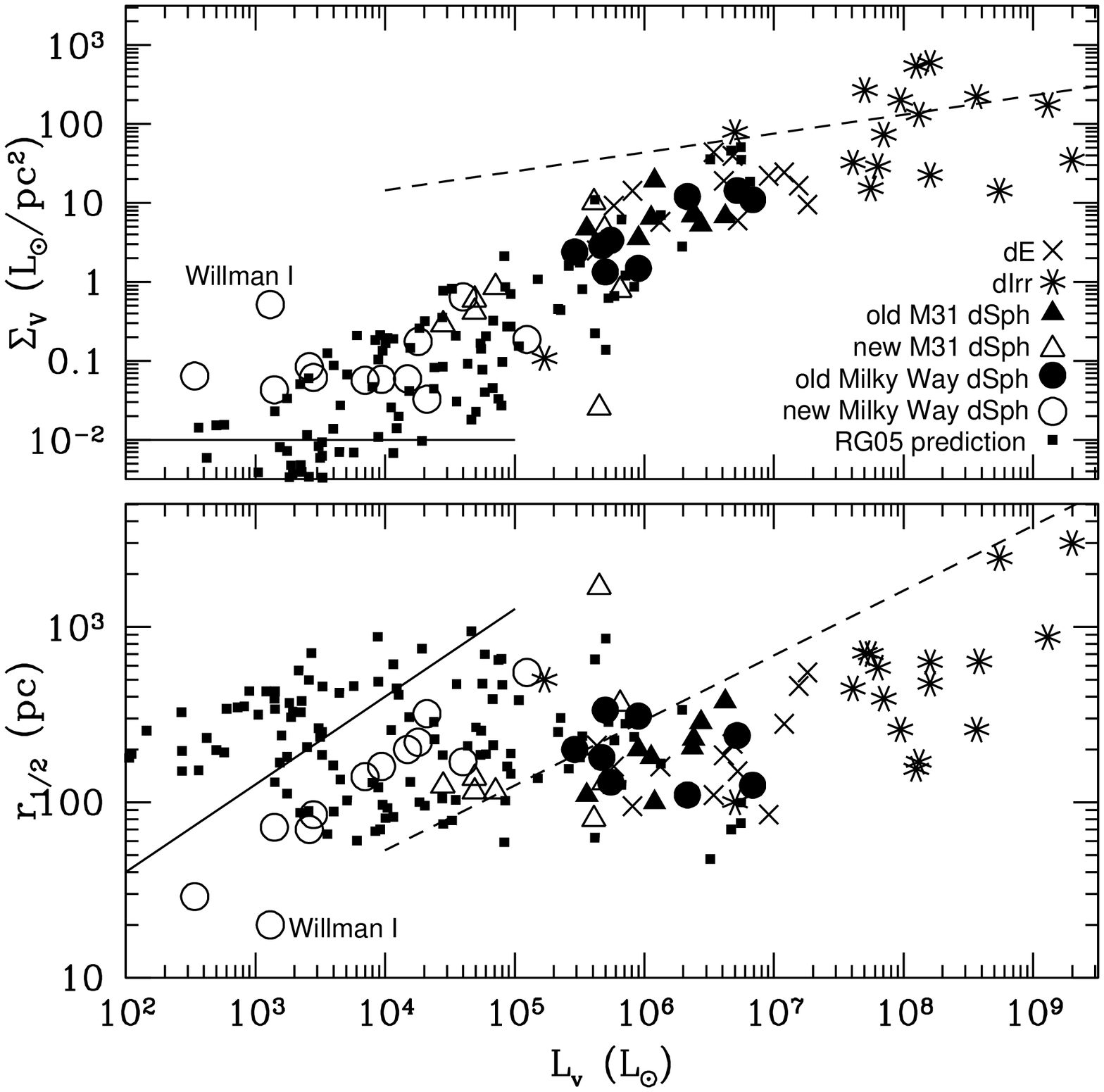}{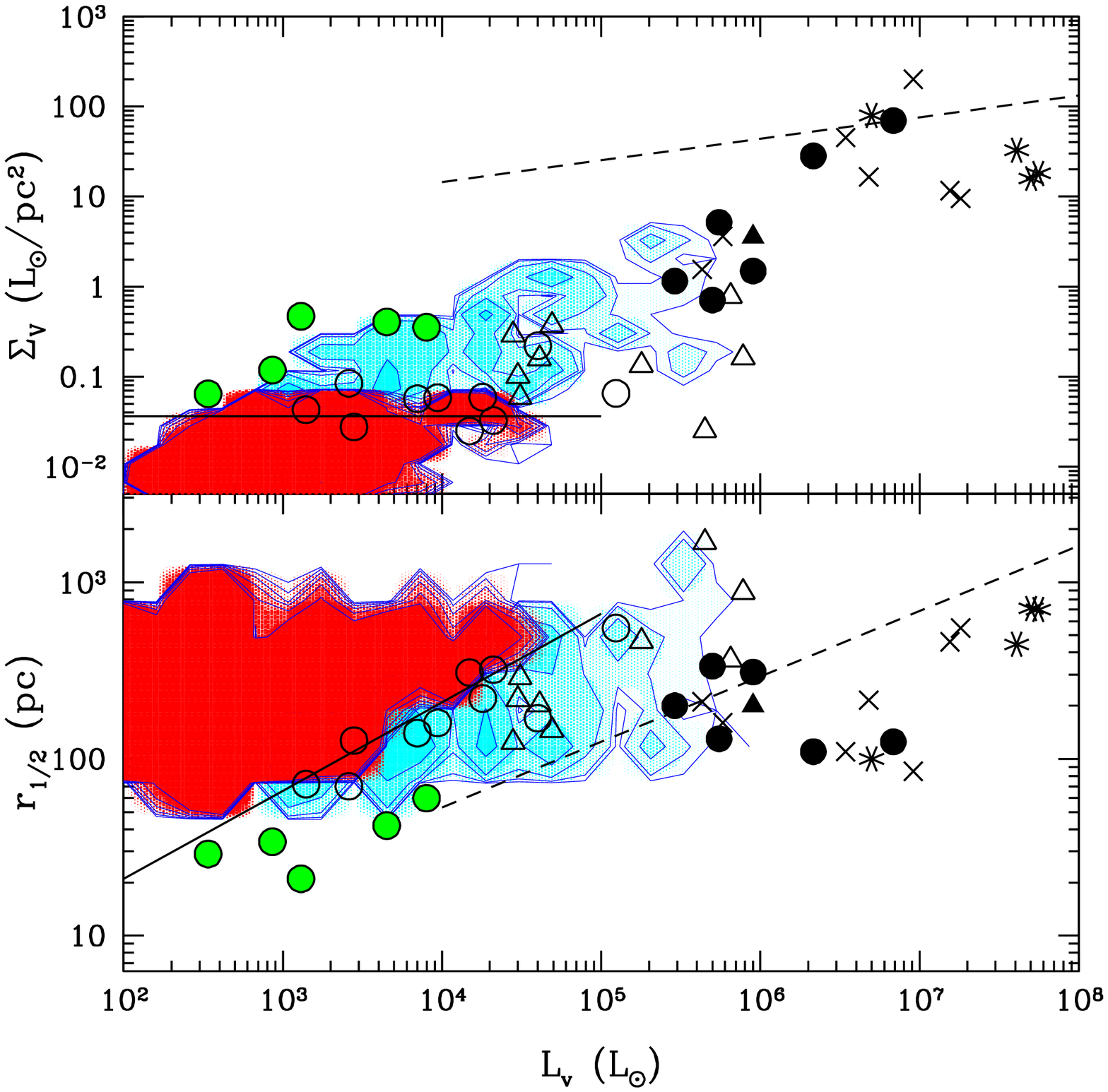}
\caption{{\it{Left}} : Figure 1 from BR09. Surface brightness and half-light radii are plotted against V-band luminosity. The small black squares show the properties of the pre-reionization halos at $z=8.3$. The other black symbols show the dwarf populations for the Milky Way and M31. The asterisks are non-fossils (dIrr), crosses are polluted fossils (dE and some dSph), the filled circles and triangles are the fossils (dSph) known before 2005 for the Milky Way and M31 respectively and the opened circles and triangles are the ultra-faint populations those galaxies found since 2005. {\it{Right}} : Surface brightness and half-light radii are plotted against V-band luminosity. The cyan contours show the distribution for the fossils from Run D and the overlaid black symbols show the observed dwarfs. In this panel we color the observed dwarfs whose half-light radii are inconsistent with our simulations green. The magenta contours show the undetectable fossils with $\Sigma_V$ below the $0^{th}$ order detection limit of the SDSS, $\sim -1.4$, \citep{Koposovetal:07}. In both panels, the solid black lines show the surface brightness limit of the Sloan \citep{Koposovetal:07} and the dashed black lines show the trends from \cite{KormendyFreeman:04} for luminous Sc-Im galaxies ($10^8 L_\odot < L_B < 10^{11} L_\odot$).}
\label{Kor.s2}
\end{figure*}

We determine the half light radii, $r_{hl}^f$, for $z=0$ fossils using the 3D $r_{hl}$ from the pre-reionization simulations, with the following assumptions. (1) The dynamical evolution of the stars is decoupled from that of the dark matter. (2) The kinetic energy of the stars is conserved. (3) The collision of the luminous pre-reionization halos is elastic with respect to the stars. (4) Enough time has passed since the collision for the halo to return to an equilibrium state. Given the kinetic energy conservation of the stars:
\begin{equation}
(\sigma_\ast^f)^2 = (L_V^f)^{-1} \sum L_V^i \times (\sigma_{\ast}^i)^2,
\end{equation}
where $\sigma_\ast^i$ and $\sigma_\ast^f$ are the 3D stellar velocity dispersions of the parent and daughter halos. For a halo in equilibrium, $r_{hl} \sim \sigma_\ast^2$, therefore:
\begin{equation}
r_{hl}^f = (L_V^f)^{-1} \sum L_V^i \times r_{hl}^i.
\end{equation}
We use $r_{hl}^f$ to calculate an average surface brightness, $<$$\Sigma_V$$>$, for our fossils in units of $L_\odot/$~pc~$^2$. The  $\Sigma_V$ and $r_{hl}$ distributions as a function of luminosity are shown in Figure~\ref{Kor.s2}.

In Figure~\ref{Kor.s2}, the black symbols are the observed Milky Way and M31 satellites overlaid on colored contours showing the equivalent distributions for the simulated true fossils. The cyan and red contours show the stellar properties of the fossils above and below the SDSS detection limits, respectively. We see that, as in BR09, our simulations are able to reproduce the observed $\Sigma_V$ and $r_{hl}$ distributions for the ultra-faint and classical dSphs, with a few exceptions. We are unable to account for the ultra-faints with $r_{hl} < 60$~pc (Coma Berenics, Segue 1 and 2, Leo V and Willman 1), all but one of which (Leo V) are within $\sim 50$~kpc of the Milky Way. 

In BR09 we called attention to an, as yet undetected, population of ultra-faints with surface brightnesses below SDSS limits. The existence of these dwarfs was independently proposed in \cite{Bullocketal:10}, who named them `stealth galaxies.' The detection of these ultra-faint dwarfs is a test for the fossil scenario. In this section, we summarize the properties expected of these extremely ultra-faint fossils.

\begin{figure*}
\plottwo{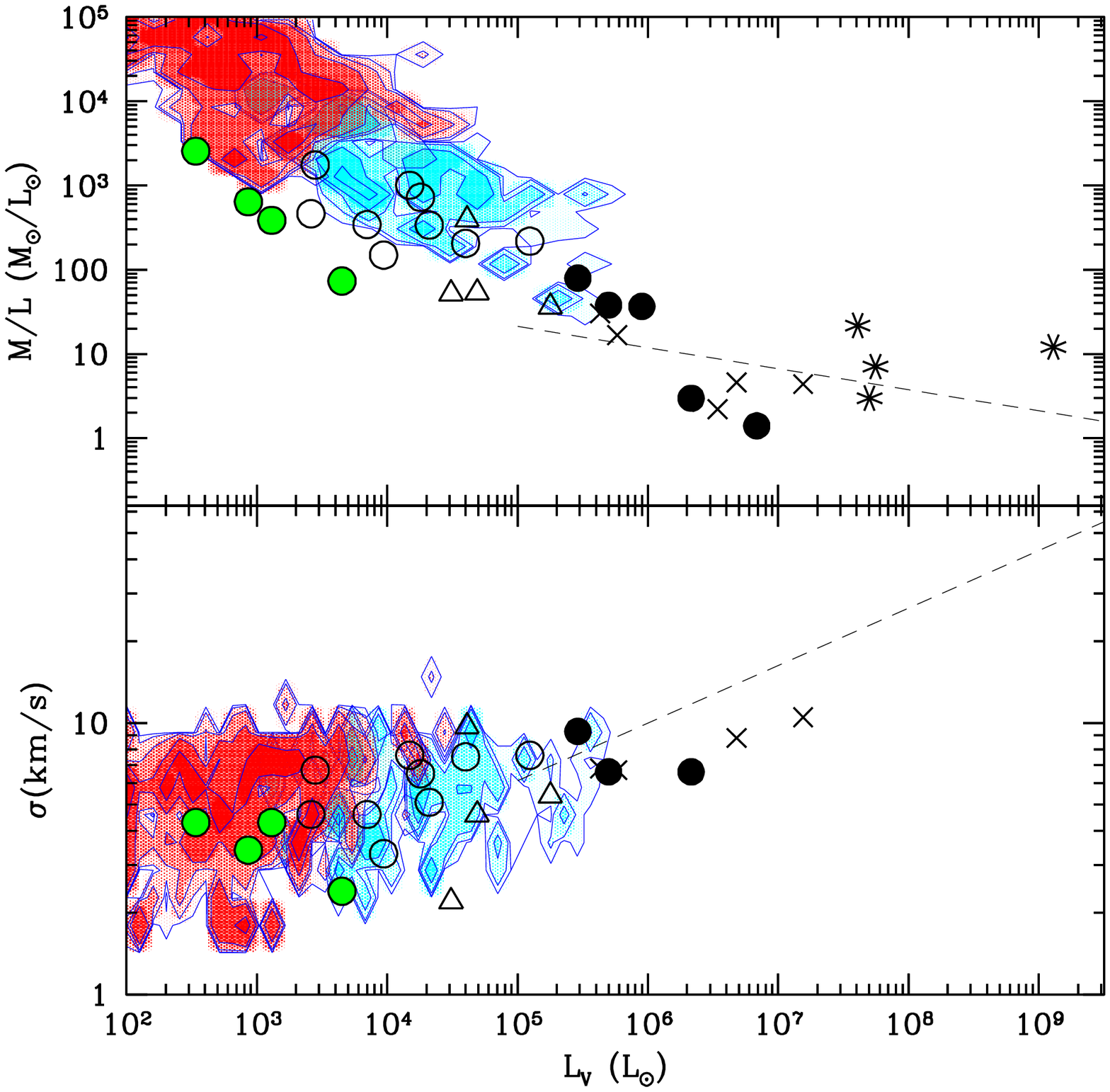}{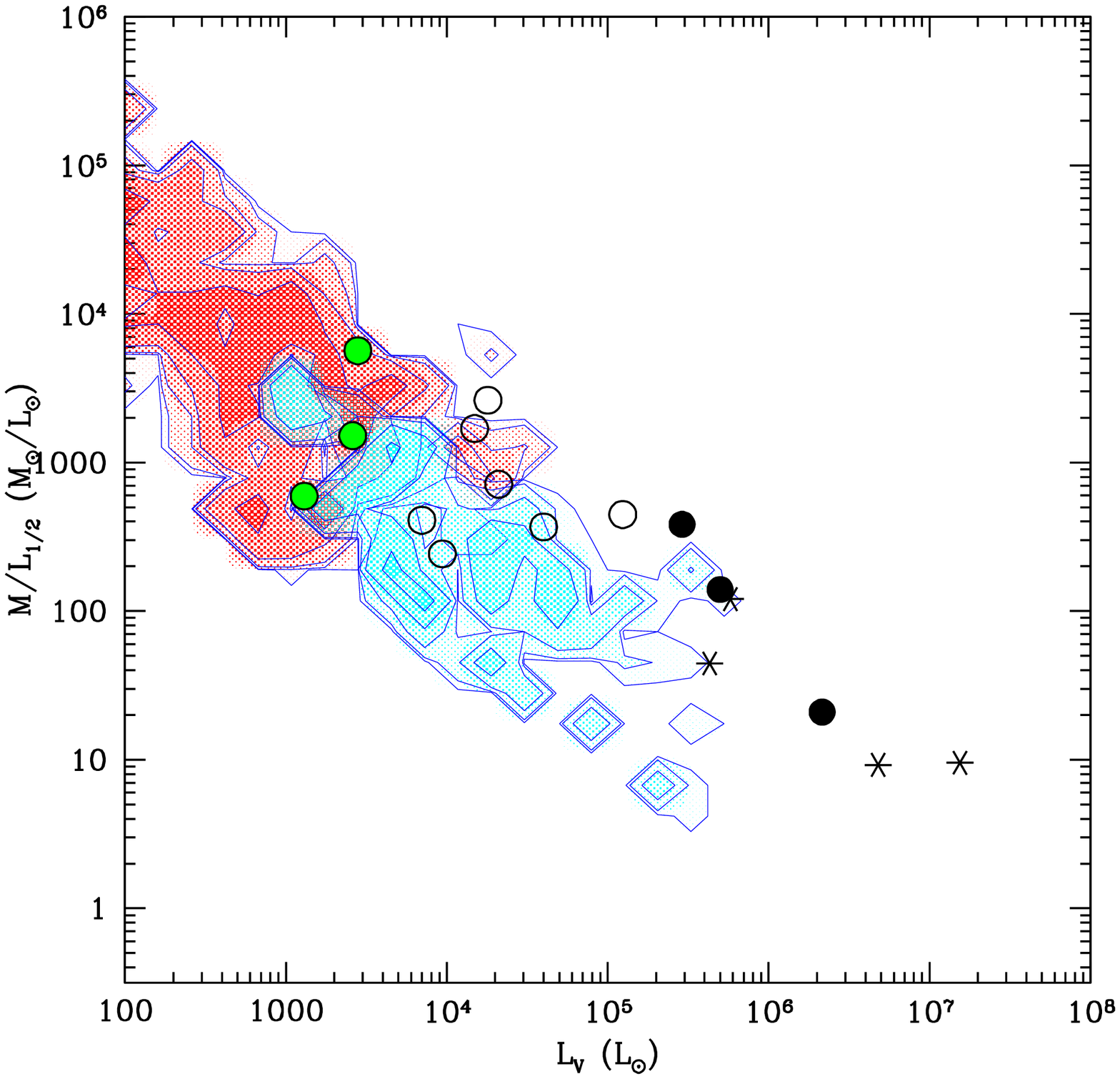}
\caption{{\it{Left}} : The stellar mass to light ratios calculated from \cite{Illingworth:76} and stellar velocity dispersions versus the V-band luminosities for Run D (blue contours) and observations (red symbols). Symbols are the same as in Figure~\ref{Kor.s2}. Once again, the dashed lines show the \cite{KormendyFreeman:04} trends for Sc-Im galaxies with $10^8 L_\odot < L_B < 10^{11} L_\odot$. {\it{Right}} : The $M/1/2L_V$ versus $1/2 L_V$ using the \cite{Wolfetal:10} mass estimator.}
\label{SigML.s2}
\end{figure*}

\begin{figure}
\centering
\plottwo{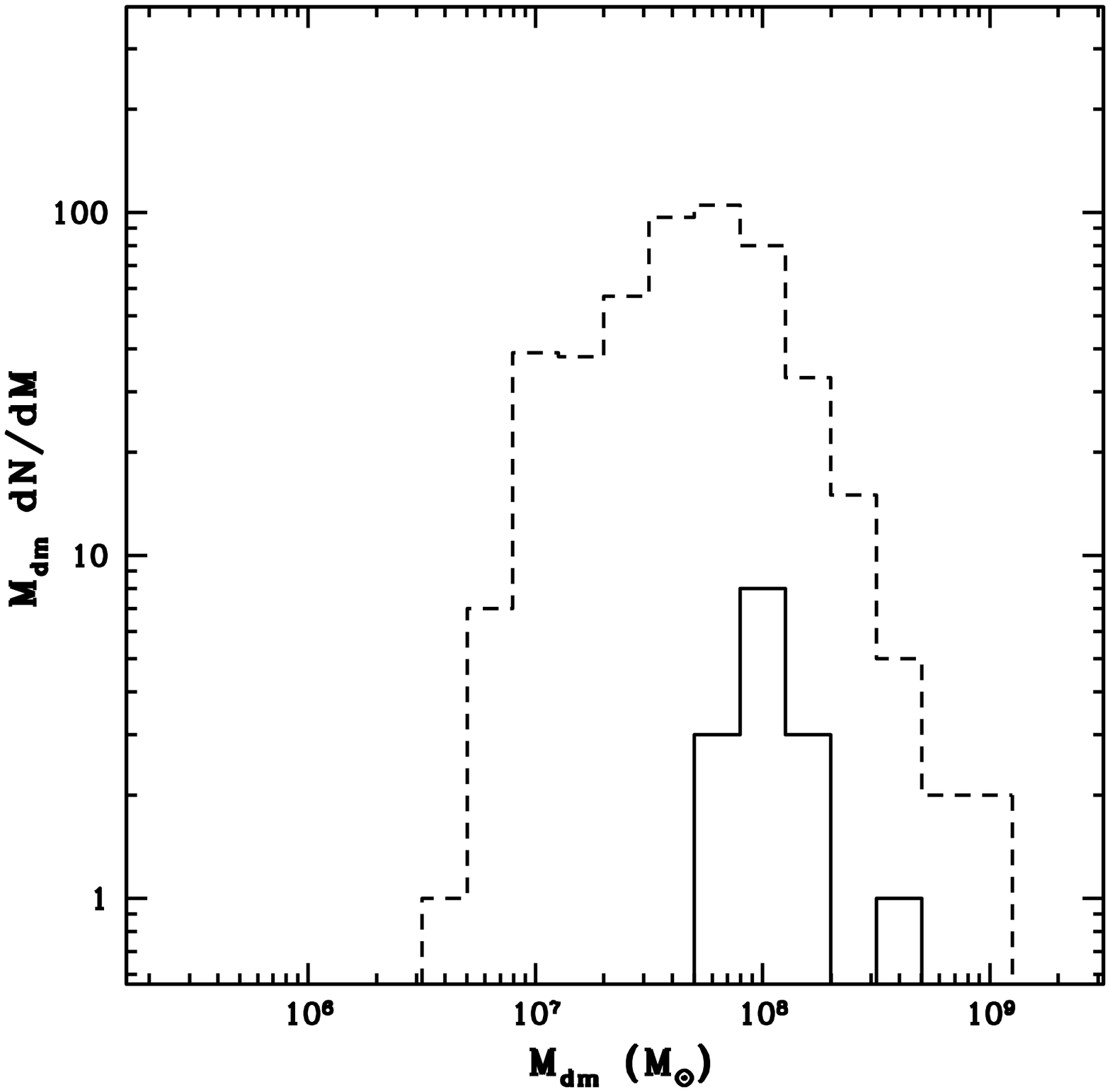}{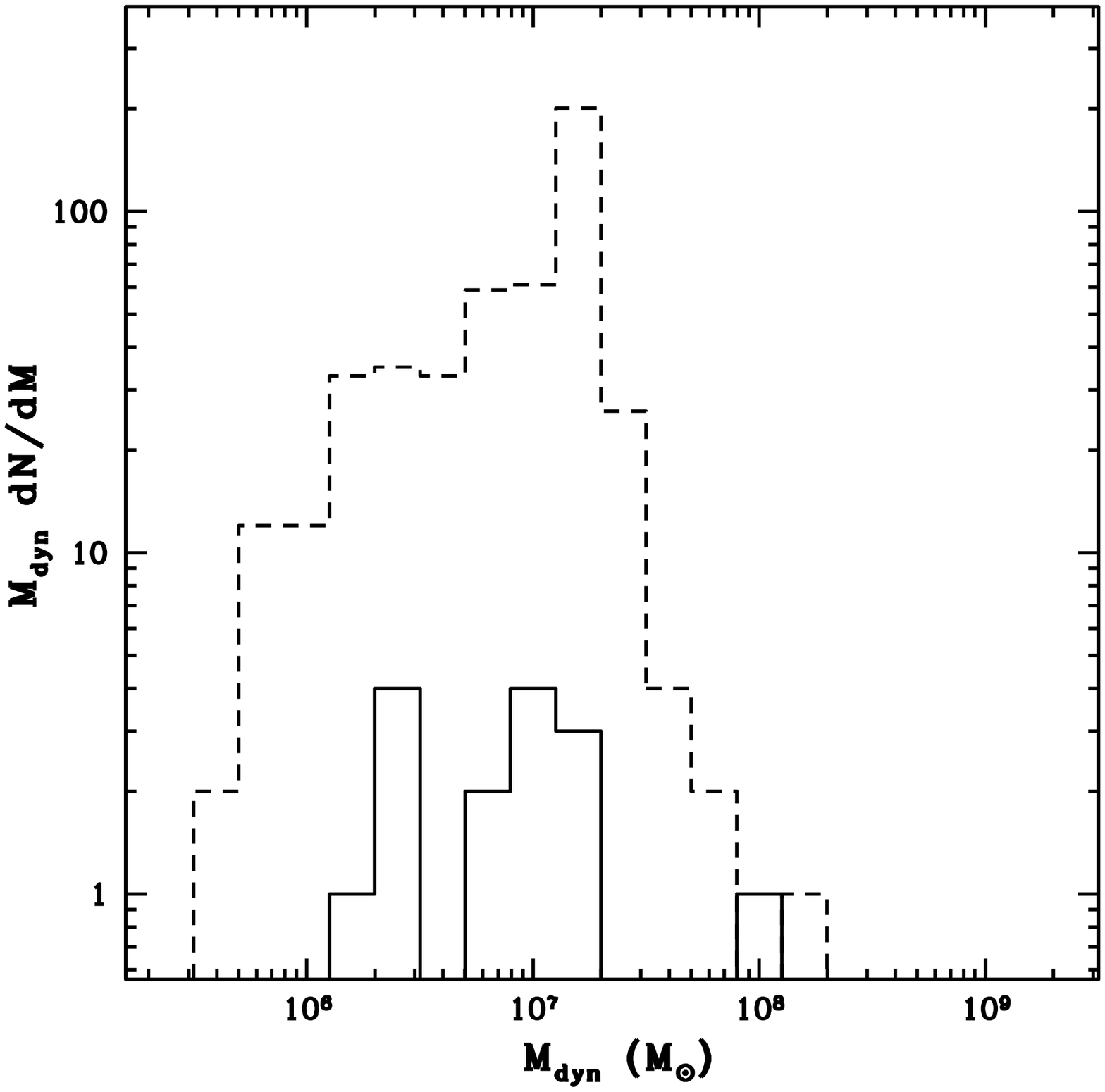}
\plottwo{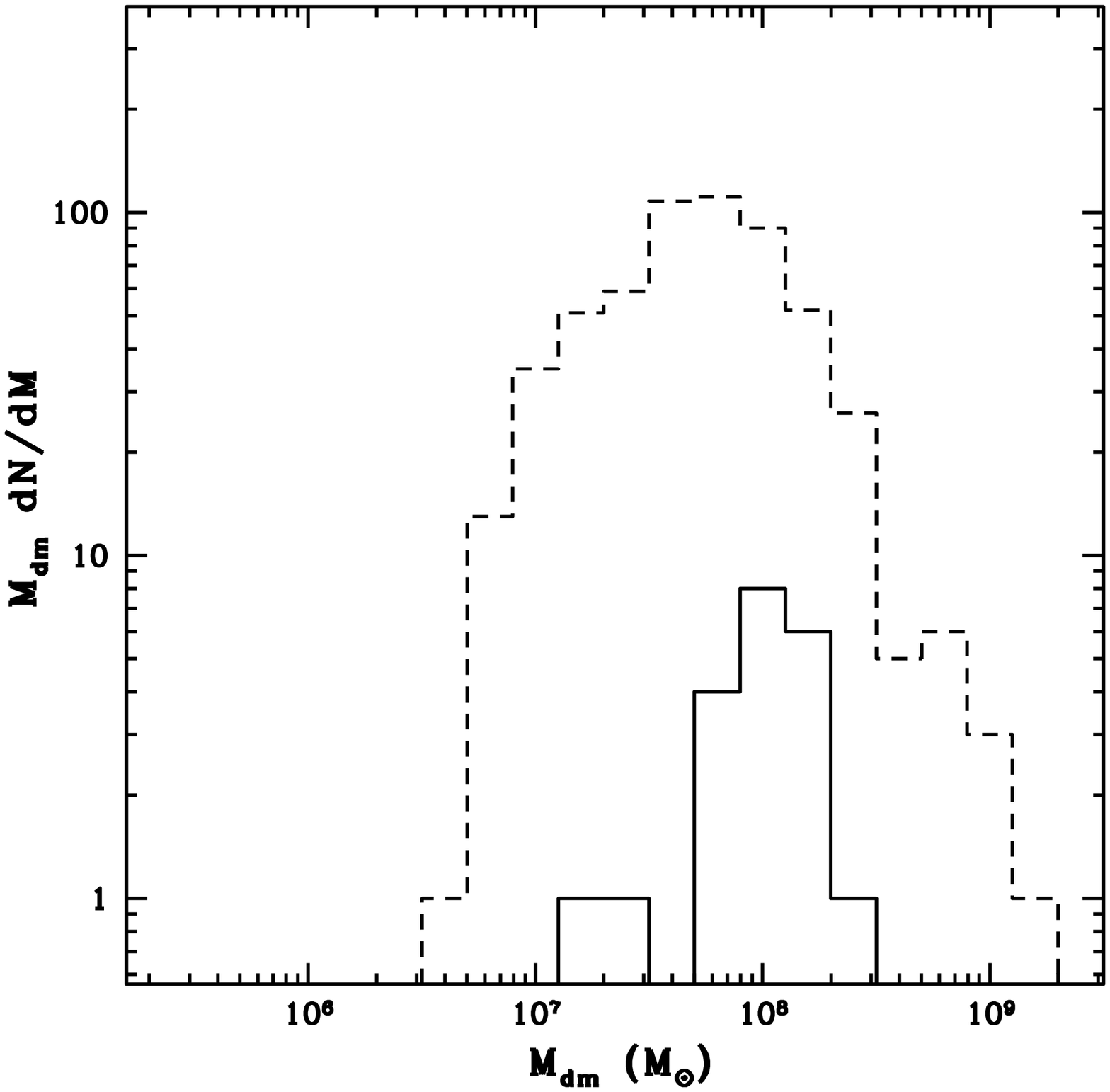}{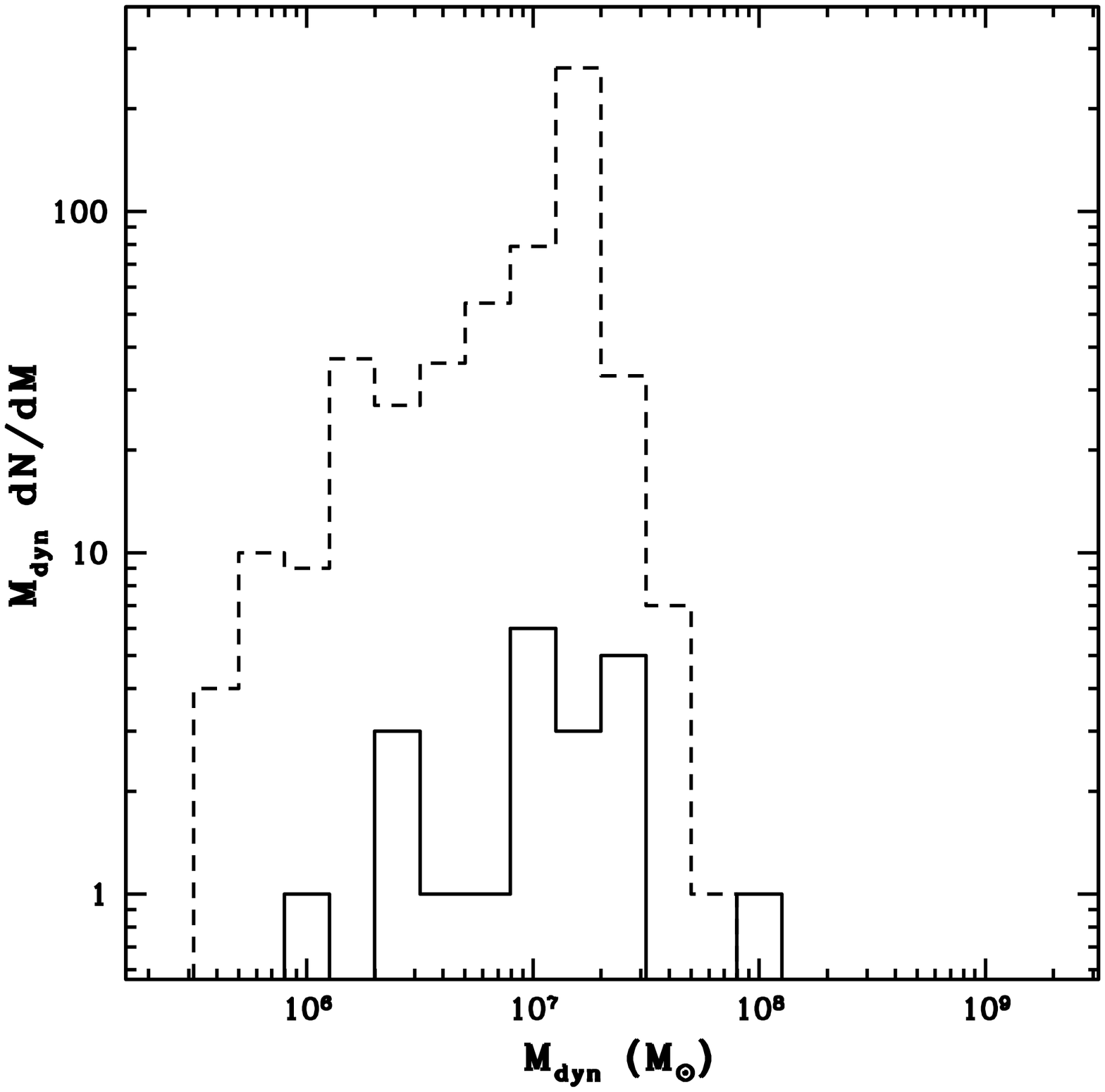}
\caption {{\it{Left}} : The mass function of the detected (solid) and undetected (dashed) fossils with $L_V>10^2 L_\odot$ within 1 Mpc of MW.2 (top) and MW.3 (bottom) from Run D. {\it{Right}} : Same as the left panels except the x-axis is the dynamical mass inside the half-light radius (Wolf et al., 2010) calculated from the velocity dispersion and half-light radius of our fossils.}
\label{MF.undetect}
\end{figure}

\begin{figure}
\plotone{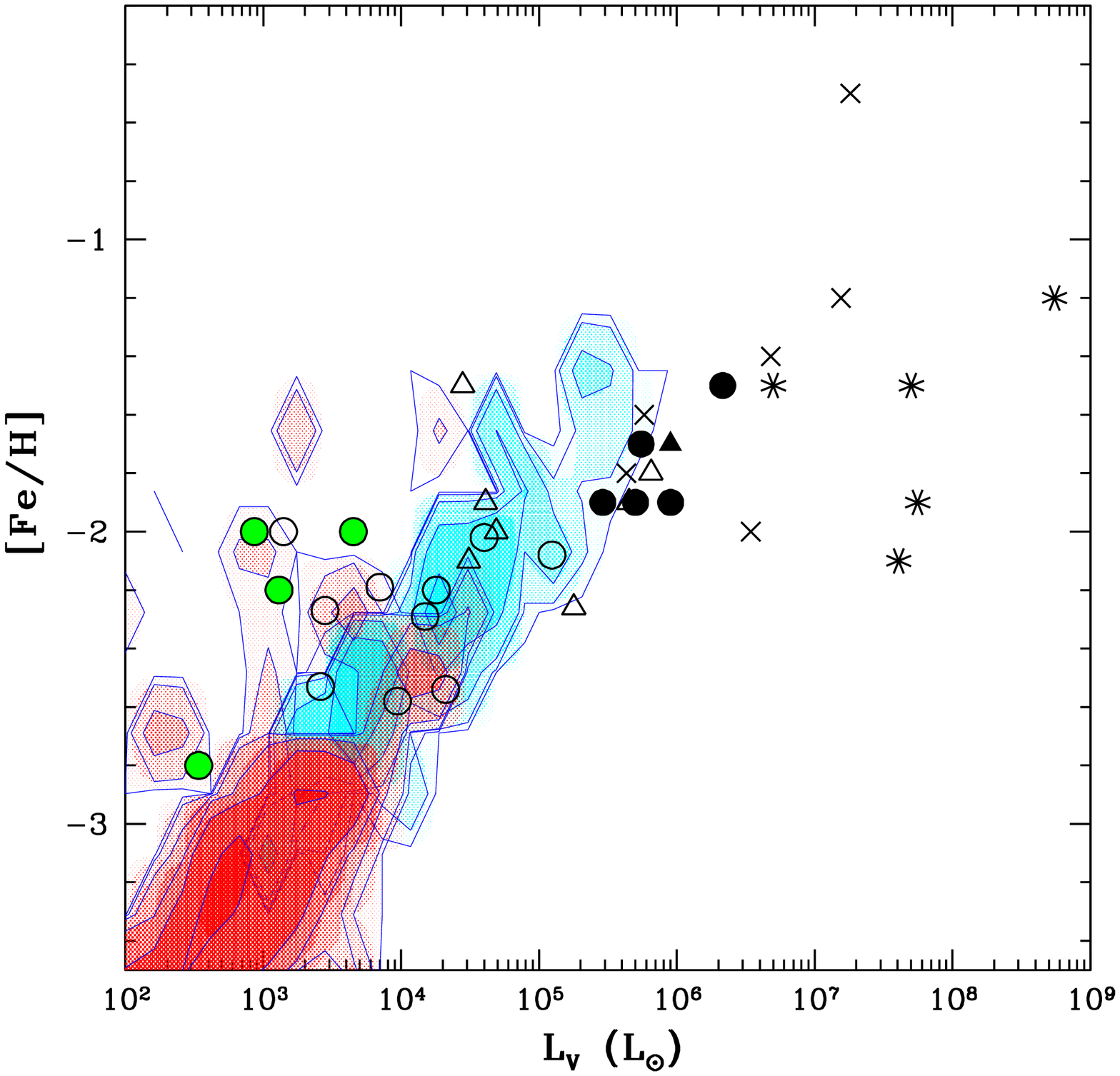}
\caption{The [Fe/H] distribution for the true fossils plotted against the V-band luminosities for Run D (blue contours) and observations (red symbols). Symbols are the same as in Figure~\ref{Kor.s2}. Our results agree with \cite{SalvadoriFerrara:09} for $L_V > 10^4$.}
\label{LZ.s2}
\end{figure}

\begin{figure}
\plotone{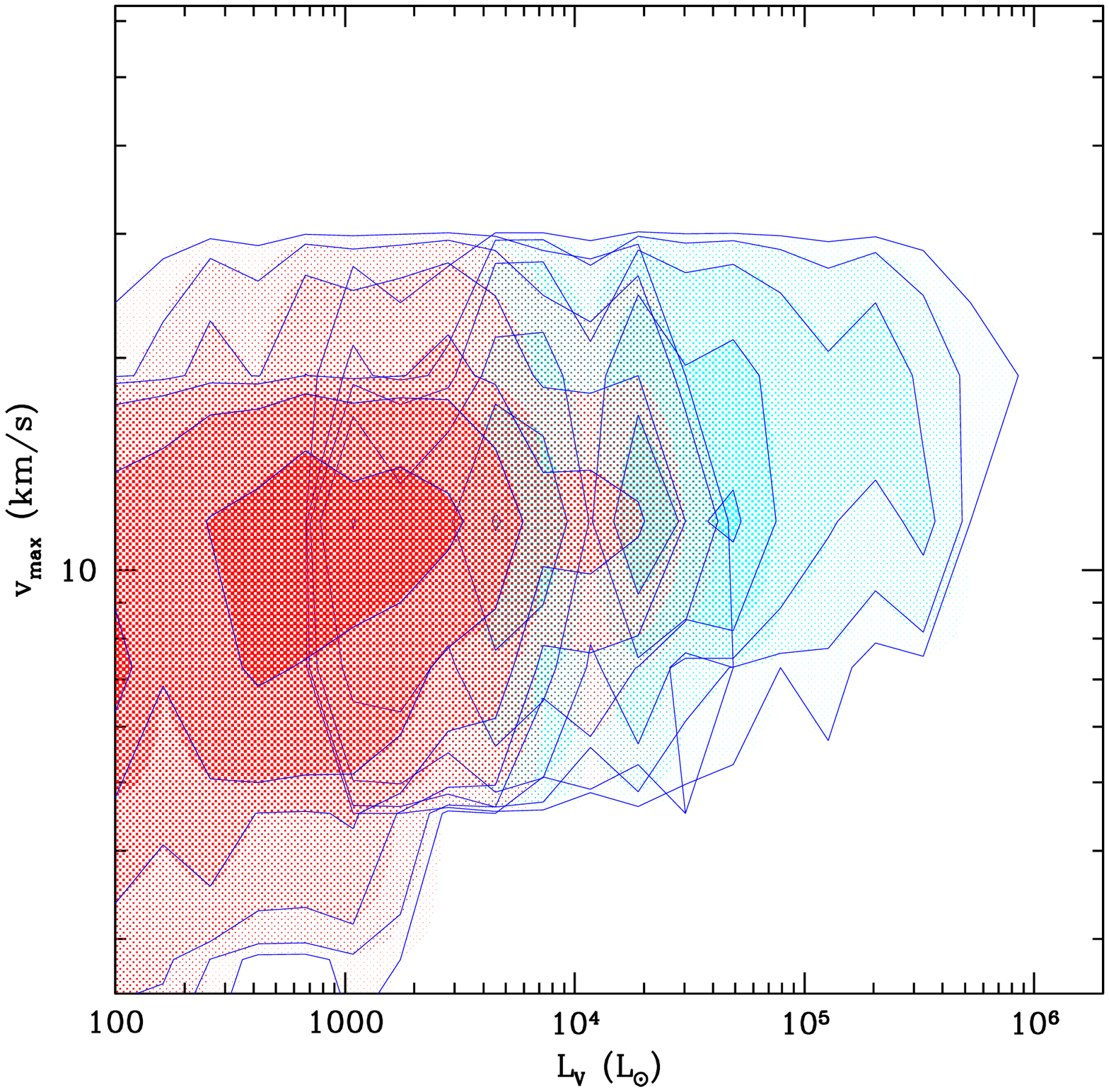}
\caption{The maximum circular velocity, $v_{max}$ of our simulated true fossils plotted against the V-band luminosities.  The cyan and red contours are the same as in Figure~\ref{Kor.s2}.  Here we show no observed dwarfs due to the lack of data.}
\label{LVmax.s2}
\end{figure}

\begin{figure}
\plotone{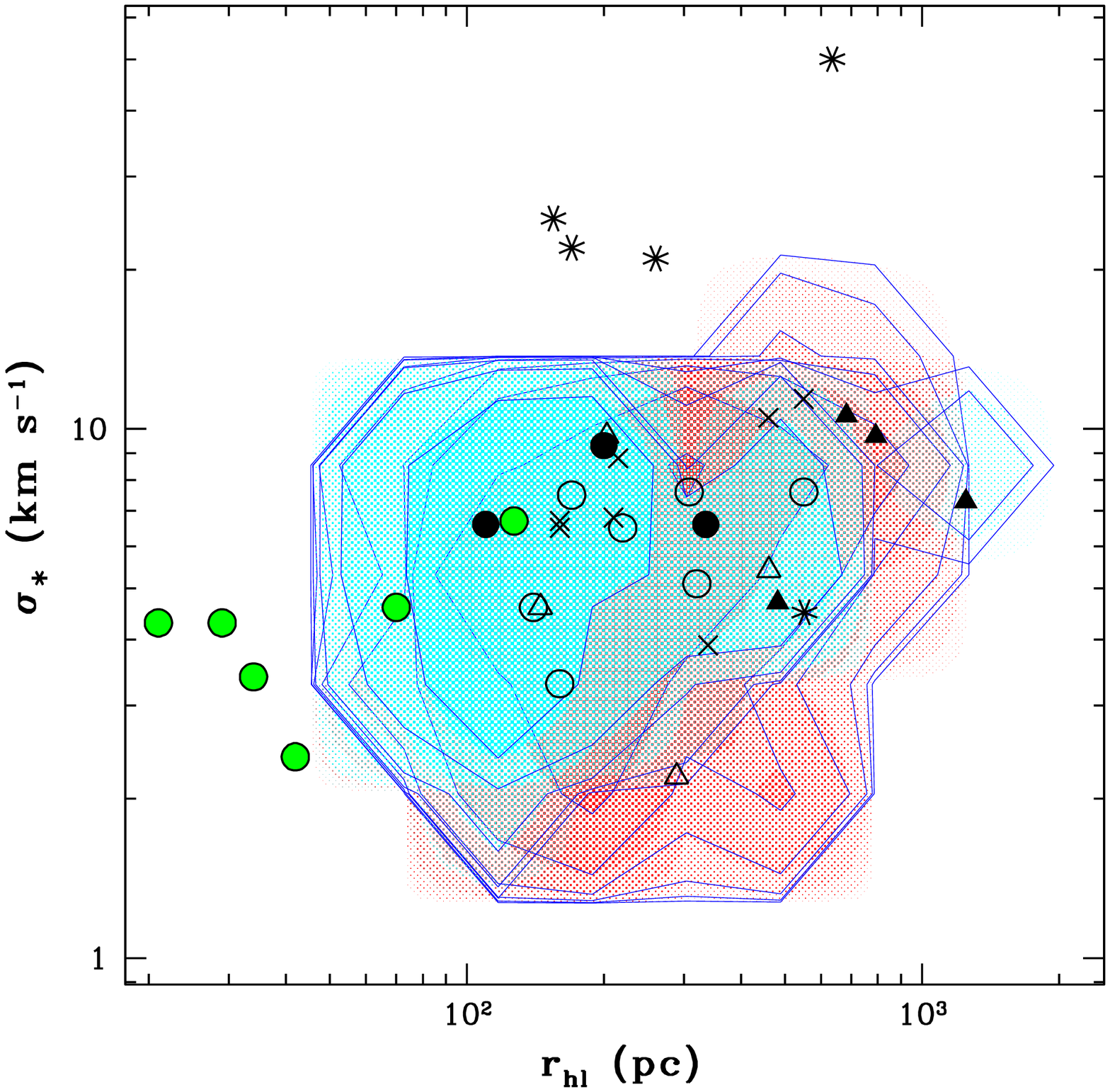}
\caption{The stellar velocity dispersion, $\sigma_\ast$ against the half-light radius, $r_{hl}$.  The black symbols are the observed dwarfs and blue and red contours from Run D have the same meanings as in Figure~\ref{Kor.s2}.}
\label{RSig.s2}
\end{figure}

In Figures~\ref{Kor.s2}~-~\ref{RSig.s2}, the simulated true fossils are shown as two sets of contours. Up until now, we have been comparing the ultra-faints and a subset of the classical dSphs to the simulated true fossils with $\Sigma_V > 10^{-1.4} L_\odot$~pc$^{-2}$. These true fossils, shown by the cyan contours, would be detectable by the SDSS \citep{Koposovetal:07}. The red contours show the true fossils which would remain undetected by SDSS. In \S~\ref{SEC.obser}, we present the existence and properties of the true fossils with surface brightnesses below the SDSS detection limits as a test for primordial star formation in minihalos. For the remainder of this section, we direct the reader to the red contours on Figures~\ref{Kor.s2}-~\ref{RSig.s2}.

The mass to light ratios and $\sigma_\ast$ of the observed and simulated populations are shown as the top and bottom of the left panel in Figure~\ref{SigML.s2}. As in Figure~\ref{Kor.s2}, the five dwarfs which do not match the $r_{hl}$ of the simulated fossils are marked with filled green circles. Excepting this subpopulation, the ultra-faints show the same distribution as the simulated true fossils for both $M/L$ and stellar velocity dispersions. In the left panel of Figure~\ref{SigML.s2}, the masses of our simulated halos are calculated from the stellar properties using \cite{Illingworth:76}. The right panel shows the mass to light ratios inside the half-light radii, $M/L_{1/2}$ versus half the V-band luminosity using the \cite{Wolfetal:10} mass estimator. The latter mass estimator is more accurate for dispersion supported systems, but we note that the agreement between the mass to light ratios of our fossils and ultra-faint dwarfs is independent of the mass estimator we use to calculate $M(\sigma_\ast,r_{hl})$. As expected, the undetected dwarfs (red contours in Figure~\ref{SigML.s2}) would have $M/L > 10^3 M_\odot/L_\odot$, higher than even the most dark matter dominated ultra-faint dwarfs. However, the range of their stellar velocity dispersion is $2-10$~km~s$^{-1}$, equivalent to the ultra-faint dwarfs and detectable fossils, and shows no evolution with decreasing luminosity.

As seen in the left panel of Figure~\ref{MF.undetect}, the mass functions of the detected and undetected fossils peak at $~10^8 M_\odot$. Note, however, that this mass function is for the total dark matter mass, not the dynamical mass calculated from the velocity dispersion and half-light radius. Our simulations provide us with the information needed to plot a mass function of the dynamical mass, refereed to in the right panel of Figure~\ref{MF.undetect} as the observed mass. For the derived observational mass function both the detected and undetected fossils peak at ~$2 \times 10^7 M_\odot$.  This peak corresponds to the `common mass scale' for dwarf spheroidals \citep{Strigarietal:08}, however, no such sharp peak is seen in the total dark matter mass function. This is because local feedback effects in small halos produce a large scatter in the relationship between total and stellar mass of halos. Two dwarfs with similar mass and extent of their stellar populations, and thus similar dynamical mass within the luminous radius may be embedded within halos whose masses vary by an order of magnitude. While the more massive halo's stellar population is concentrated at the center of its potential, the lower mass halo's stars fill a larger fraction of its dark matter halo. This produces either two halos with the same dark matter mass and different $M_{300}(r_{hl},\sigma_\ast)$ or vice versa, a common $M_{300}$ but very different dark matter masses.

For the metallicity distribution, we also use a luminosity weighted average:
\begin{equation}
[Fe/H]^f = log(\sum 10^{[Fe/H]^i} \times L_V^i) - log(L_V^f).
\end{equation}
The distribution of metallicity versus $L_V$ is shown for our $z=0$ fossils and the known ultra-faint and classical dwarfs.  As in BR09, the fossil metallicities from Run D are consistent with the observed distribution for the ultra-faint and classical dSph. We also find out results for $L_V>10^4 L_\odot$ to be in agreement with \cite{SalvadoriFerrara:09} while for the dimmest fossils our work finds comparatively lower metallicities. The undetected dwarfs (red contours in Figure~\ref{LZ.s2}) have $[Fe/H]<-2.5$ and as low as $-3.5$ with slightly larger scatter than their detectable counterparts.

The maximum circular velocity versus $L_V$ contours for our simulated true fossils are shown in Figure~\ref{LVmax.s2} to illustrate the following. While $v_{max}$  does decrease by approximately a factor of two over four decades of luminosity, the scatter in $v_{max}$ at a given $L_V$ is large. Though a halo with $v_{max} < 6$~km~s$^{-1}$ is likely to have a $L_V < 10^4 L_\odot$, there is, at most, a minimal trend of decreasing $v_{max}$ with decreasing luminosity for the primordial fossils. This highlights a theme across all our stellar property comparisons. Because of the strong dependence of their stellar properties on stochastic feedback effects, there is no baryonic property that shows a strong trend with maximum circular velocity and the size of the dark matter halo.

We now briefly discuss the M31 satellite population. Figure~\ref{RSig.s2} shows the $\sigma_\ast$ plotted against $r_{hl}$ on a similar scale to the top left panel of Figure 18 in \cite{Collinsetal:10}. The circles show the Milky Way dSphs, while the triangles show the dSphs associated with M31. We find that four of the six M31 dSphs plotted are within, abet at the edges of, the contours of detectable true fossils. Like their Milky Way counterparts, the new M31 dSphs show reasonable agreement with our simulated primordial fossils excepting of $r_{hl}$, which are higher than expected by our simulations for two of the M31 dwarfs. However, this does not represent a major problem for our model since $\sim65\%$ of simulated true fossils with $L_V > 10^4 L_\odot$ have undergone one or more major mergers that may have puffed up their stellar populations. Our estimates do not account for extra heating of the stellar populations by the kinetic energy of the collision. A higher $\sigma_\ast$ would result in a more extended stellar population in the same mass halo. We will discuss the comparison between the M31 dSphs and our simulated fossil dwarfs in an upcoming paper.

\subsection{The Inner Ultra-Faints}
\label{SEC.tidal}

In this section, we discuss a possible origin scenario for the inner ultra-faint dwarfs, ie. the ultra-faints whose half light radii and mass to light ratios are lower than our true fossils. These dwarfs are, Segue 1 and 2, Leo V, Pisces II and Willman 1, and excepting Leo V and Pisces II (both at $\sim180$~kpc) all are within $50$~kpc of the Milky Way.

However, their mass to light ratios follow a shifted power law with a similar slope to the true fossils and more luminous dwarfs. The stellar velocity dispersions are in the range expected for primordial fossils, but the inner ultra-faints show an $L_V-\sigma_\ast$ combination which would be expected for true fossils below the detection limits of SDSS (red contours on Figure~\ref{SigML.s2}).  These properties are either directly affected by tidal stripping ($r_{hl}$ and $\sigma_\ast$) or are derived from affected properties ($\Sigma_V$ and $M/L$).  However, the metallicity of the stars is not affected by tidal stripping.

Figure~\ref{LZ.s2} shows the metallicities of the inner ultra-faint dwarfs do not fall on the luminosity-metallicity relation. However, their scatter is consistent with expectations for true fossils. To place the Segues, Leo V, Pisces II and Willman 1 on the luminosity-metallicity relation traced by the majority of the ultra-faints and our fossils, their luminosities would need to be increased by one to two orders of magnitude. We suggest these dwarfs may be a subset of bright primordial fossils which have been stripped of $90\%-99\%$ of their stars. 

\begin{table}
\begin{centering}
\begin{tabular}{ c || c | c }
\hline
& $R < 50$~kpc & $R > 50$~kpc \\
\hline
\hline
\multirow{3}{*}{Inconsistent} & Segue 1 & Pisces II $^\ast$ \\
& Segue 2 & Leo V $^{\ast}$\\
& Willman 1 & \\
\hline
\multirow{5}{*}{Consistent} & \multirow{5}{*}{Coma Ber.} & Bootes I $\&$ II \\
& & CVn I $\&$ II \\
& & Hercules \\
& & Leo IV \& Leo T \\
& & Ursa Major I \\
\hline
\end{tabular}
\caption[Inner Ultra-faint dwarfs as tidal ultra-faint dwarfs?]{Table of Milky Way ultra-faint dwarfs classified by their distance from our galaxy (columns) and whether or not they are consistent with our predictions for the fossils of the first galaxies (row). Note the correlation between distance and consistency. (*) Pisces II and Leo V are both on the lower end of radii expected for fossils, as such they are marked as ``inconsistent,'' but are not as far from predictions as the ``inconsistent'' ultra-faints within $50$~kpc.}
\end{centering}
\end{table}

\section{Observational Tests}
\label{SEC.obser}

In this section, we present a set of observational tests which can provide support for a primordial formation scenario for the faintest Milky Way satellites.
 
Better determination of whether the ultra-faints are being tidally disrupted can help determine whether a subset of the faintest Milky Way satellites are pristine fossils. The ultra-faint dwarfs whose $r_{hl}$ do not match our simulated true fossils, shown as filled green circles in Figures~\ref{Kor.s2}~-~\ref{RSig.s2}, display the signs of being tidally disrupted by the Milky Way, including proximity to the Milky Way ($R<50$~kpc). While Willman 1, Segue 2, Leo V, and Coma Berenticis show signs of tidal disruption, this does not prove the primordial scenario. However, it would place their origin as disrupted objects in line with our proposal in Section~\ref{SEC.tidal}. If additional observations show these tidal ultra-faints are ${\it{not}}$ tidally disrupted, then our primordial formation model can not explain their current properties. The exception to this picture is Segue 1. The tidal status of Segue 1 has been recently debated \citep{Niederste-Ostholtetal:09,Norrisetal:10,Martinezetal:10}, and \cite{Simonetal:10} show that the kinematics of the stars are consistent with a stellar population well within the tidal radius. The high density of the stars suggests that Segue 1 formed at high redshift in a rare, high $\sigma$ peak which our $1$~Mpc$^{3}$ pre-reionization volume is not large enough to include.

The number of Milky Way satellites alone provides a test for star formation in minihalos. For a given filtering velocity, there is a number of satellites, $N_{nf}$, which has a $v_{max}$ at infall above the filtering velocity. For $v_{filt}=20$~km~s$^{-1}$, $N_{nf}$ is $90\pm10$ and for $v_{filt}=30$~km~s$^{-1}$, $N_{nf}$ is $60\pm8$, the latter is equivalent to the number of currently known Milky Way satellites. If the number of satellites, $N_{sat}$, is greater than $N_{nf}$, some minihalos had to have formed stars before reionization. Conversely, if $N_{sat}<N_{nf}$, either no minihalos formed stars or none survived near the Milky Way.

We next outline the stellar properties we can expect of the undetected dwarfs around the Milky Way if they are part of a population of fossils of the first galaxies. The red contours of Figures~\ref{Kor.s2}~-~\ref{RSig.s2} show the properties of the predicted population.

\begin{enumerate}

\item {\it{Half-light radii}}

The undetected dwarfs should have the same distribution of half-light radii as the currently known, ultra-faint population, from $\sim100$~pc to $\sim1000$~pc.

\item {\it{Mass to Light Ratio}}

The mass to light ratio of undetected dwarfs should generally be greater than $10^3 M_\odot/L_\odot$ and as high as approximately $10^5 M_\odot/L_\odot$ and follow a roughly linear relation for dwarfs with $L_V < 10^5 L_\odot$.

\item {\it{Stellar velocity dispersion}}

There should be no decrease in the stellar velocity dispersion, $\sigma_\ast$, with V-band luminosity. This directly contradicts the decreasing $\sigma_\ast$ with $L_V$ seen for tidally stripped dwarfs in Figure 4 of \cite{WadepuhlSpringel:10}.

\item {\it{Metallicity}}

The undetected dwarfs should have typical $[Fe/H]<-2.5$, a significant number with $[Fe/H]<-3.0$. This would make these undetected dwarfs excellent candidates for the search for ultra-metal poor stars \citep{Frebeletal:10,FrebelB:10}.

\end{enumerate}

\section{Summary and Conclusions}\label{summary}

In this paper, the first of a series, we present a new method for generating initial conditions for cosmological N-body simulations which allows us to create simulated maps of the present-day distribution of fossils in a ``Local Volume.'' In order to produce these maps, we assume pre-reionization fossils do not accrete gas and form stars after reionization. They are hosted in dark halos that maintain circular velocities below a critical threshold, $v_{filt} \sim 20-30$~km/s. The precise value of $v_{filt}$ depends on details of the reheating in the local IGM by stars and AGN. Therefore, we explore values different values for $v_{filt}$. For our purposes, we do not need to include gas dynamics. The lack of post-reionization baryonic evolution in the fossils allows us to simply simulate the evolution of the dark matter and stars using N-body techniques. 

We have combined the results from previous cosmological simulations of the formation of the first galaxies \citep{RicottiGnedinShull:02b, RicottiGnedin:05, RicottiGnedinShull:08} with N-body simulations in which each particle in the initial conditions represents a pre-reionization minihalo.  Our N-body simulations zoom in on a Local Volume containing one to two Milky Ways. We follow the merger history and tidal stripping of pre-reionization fossils as they merge to form more massive galactic satellites of the Milky Way. We also trace the evolution of more massive non-fossil satellites, but we do not account for star formation taking place after reionization.

Our goal is to determine if a widespread population of primordial dwarfs is consistent with the observed population of Milky Way and Andromeda satellites, and, at the same time, matches observations of dwarfs in the Local Void. It is not well established whether halos with $v_{max}<20$~km/s, too small to initiate collapse via Lyman-alpha cooling, remain dark or form luminous dwarfs. Our simulations are a first attempt to constrain the theory of self-regulated galaxy formation before reionization using ``near field'' observations. Observational tests based on our results can constrain models of star formation in minihalos before reionization.

In this paper, we present and validate our method by comparing our results with published cosmological N-body simulations. We then present maps of the Local Volume showing the distribution of stars formed before reionization in the present day universe. We find that primordial fossils are present in the voids regardless of the details of reionization, however, reionization by X-rays produces darker voids. Finally, we show plots of the present-day properties of true-fossils and compare them to observations of classical and ultra-faint dwarfs for the Milky Way and Andromeda. We show that the simulated properties of fossils agree with a subset of ultra-faint dwarfs discovered in the Milky Way and Andromeda, and reiterate the BR09 result of a large population of fossils with surface brightness below the SDSS detection limits. The properties of this ``stealth'' population are shown as red contours in Figures~\ref{Kor.s2}~-~\ref{RSig.s2}.

We find that most classical dSph satellites are unlikely true-fossils of the first galaxies, even though they have properties expected of fossils: diffuse, old stellar populations with no gas \citep{RicottiGnedin:05, BovillRicotti:09}. The reason that true-fossils in the Milky Way have luminosities $<10^6$~L$_\odot$, is that the most luminous pre-reionization fossils, with $v_{max} \sim 20$~km/s form in over-dense regions and are strongly clustered. Thus, they are likely to merge into more massive non-fossil dwarfs. The surviving fossils found today are a sub-population with lower typical luminosities, and form in less clustered regions in which feedback effects suppress rather than stimulate star formation.

Other results found in this first paper are highlighted below.
\begin{enumerate}

\item Voids contain many low luminosity fossil galaxies. However they have surface brightnesses and luminosities making them undetectable by SDSS. One possible way to detect these void dwarfs is if they experience a late phase of gas condensation from the IGM as proposed in \citep{Ricotti:09}. Future and present 21cm surveys such as ALFALFA and GALFA may be used to find these objects \citep{Giovanellietal:05, Begumetal:10}.

\item We find a linear scaling relation between the number of luminous satellites and the mass of host halos. The scaling has scatter similar to the relationship between the total number of sub-halos with $M>10^7$~M$_\odot$ ($v_{max}>5$~km~s$^{-1}$) and the host mass, although the normalization is $3-4$ times lower.

\item Overall $\sim 25\%$ (for $v_{filt}=20$~km~s$^{-1}$) to $\sim30\%$ (for $v_{filt}=30$~km~s$^{-1}$) of the primordial fossils at the present day have undergone a merger with another luminous fossil. This fraction increases with the modern luminosity of the dwarf. Hence, the typical half light radii of this population can be larger than the original distribution at reionization. These fossils are even harder to detect due to their lower surface brightness. This effect also increases the spread of the relationship between half light radii vs. luminosity and surface brightness vs. luminosity of fossils at $z=0$.

\item Due to the dependence of the properties of primordial dwarfs on their formation environment \citep{RicottiGnedinShull:08}, we find very few true fossils with $L_V>10^6 L_\odot$, and none within $1$~Mpc of our Milky Ways. This places the identification of some of the more luminous classical dSphs fossils in doubt.

\item Leo V, Pisces II, Segue 2 and Willman 1 have half-light radii which are too small, and metallicities too large, for their luminosities. Due to their proximity to the Milky Way, we speculate that their stars and dark halos have been affected by tides.  Hence, these ultra-faints may represent a population of massive primordial dwarfs which have lost $\simgt90\%$ of their stars via tidal interactions.

\item We reiterate the existence of a yet undetected population of fossils with luminosities $L_V < 10^4$~L$_\odot$ and surface brightness $<10^{-1.4} L_\odot$~/pc$^2$~. We present plots showing, in detail, the expected properties of this population. We also notice that some of the new ultra-faint satellites in Andromeda have half light radii in agreement with the properties of the undetected fossil population, but luminosities $\sim 10^5 L_\odot$.

\end{enumerate}

The second paper of this series (Bovill \& Ricotti, 2010b), will focus on studying the galactocentric radial distribution and cumulative luminosity functions of simulated fossil and non-fossil satellites and compare the results to the Milky Way dwarfs. Although we have not unmistakably demonstrated the existence of fossils around the Milky Way using existing observations, we propose new observational tests already feasible with current instruments that should tell us whether pre-reionization fossil formation was widespread in the early universe.

In a future paper, we will run simulations in which we follow each star and dark matter particle in our pre-reionization simulations to $z=0$. These higher resolution simulations will allow us to study tidal stripping of stars and dark matter for halos within $50$~kpc from the Milky Way and better understand the nature of the faintest ultra-faint dwarfs. This method will also allow us to better quantify the stellar properties of non-fossil dwarf satellites formed by mergers of fossils.

\acknowledgements

The simulations presented in this paper were carried out using computing clusters administered by the Center for Theory and Computation of the Department of Astronomy at the University of Maryland ("yorp"), and the Office of Information Technology at the University of Maryland ("hpcc"). This research was supported by NASA grants NNX07AH10G and NNX10AH10G. The authors thank the anonymous referee for constructive comments and feedback. MSB and MR would also like to thank Stacy McGaugh, Susan Lamb, Rosie Wyse, and Derek Richardson for helpful conversations and comments on this work.

\appendix

\section{First Order Initial Conditions}
\label{FO}

To study the distribution and stellar properties of the fossils of the first galaxies across a region equivalent to the Local Volume, we need an N-body simulation with the dynamical range to resolve $M<10^7 M_\odot$ halos through a volume $\sim10$~Mpc on a side. We also want to directly trace the present day distribution of the halos from our pre-reionization simulations, while keeping the required computational resources reasonable.

We generate hybrid initial conditions in which the distribution of matter is governed by different mechanisms on different scales. As in traditional N-body simulations, structure on scales $l>1$~Mpc, is set by a power spectrum. However, on scales $l<1$~Mpc, the positions, velocities and masses of the particles are set by the Press-Schechter, via outputs from our pre-reionization simulations \citep{RicottiGnedinShull:02a,RicottiGnedinShull:02b,RicottiGnedinShull:08}. These hybrid initial conditions are set as follows. (1) We locate an analog to the Local Volume within a coarse resolution $50^3$~Mpc$^{3}$ volume run from $z=40$ to $z=0$. (2) A high resolution region is built out of the final outputs from the pre-reionization simulations at $z_{init}=8.3$. (3) Finally, we insert our high resolution, `Local Volume' into the larger coarse resolution simulation at $z_{init}=8.3$ and run it to $z=0$ using Gadget 2 \citep{Springel:05}.  In the rest of Section~\ref{FO} we will explain these steps in more detail.

We need to generate and run a volume large enough to contain at least one subvolume analogous to the Local Volume. Our coarse resolution simulation is a $50^3$~Mpc$^3$ volume with $250^3$ particles run from $z=40$ to $z=0$. The power spectrum at $z=40$ is generated by P-GenIC. At $z=0$, we use HOP \cite{EisensteinHut:98} to locate potential Milky Ways. In our `Local Volume,' we look for a filament between two Virgo-like clusters with $2-3$ halos with $M \sim 10^{12} M_\odot$ within a $7^3-10^3$~Mpc$^3$ volume. Ideally, one of our Milky Ways has an equal mass companion within $1$~Mpc, however, we were not able to find such a pair in our volume.

From the location of our Milky Ways, we define our `Local Volume' as a region $\sim5-10$~Mpc across, centered on one of our Milky Ways. Once we have a `Local Volume' at $z=0$, we estimate its equivalent volume at $z_{init}=8.3$. We do this via tagging the coarse resolution particles in our present day Local Volume, and, using their positions at $z=8.3$, determine the equivalent rectangular prism containing the majority of the tagged particles. At this point, we have defined a high resolution region with dimensions $m \times n \times p$.

\begin{figure}
\centering
\includegraphics[height=150mm,width=150mm]{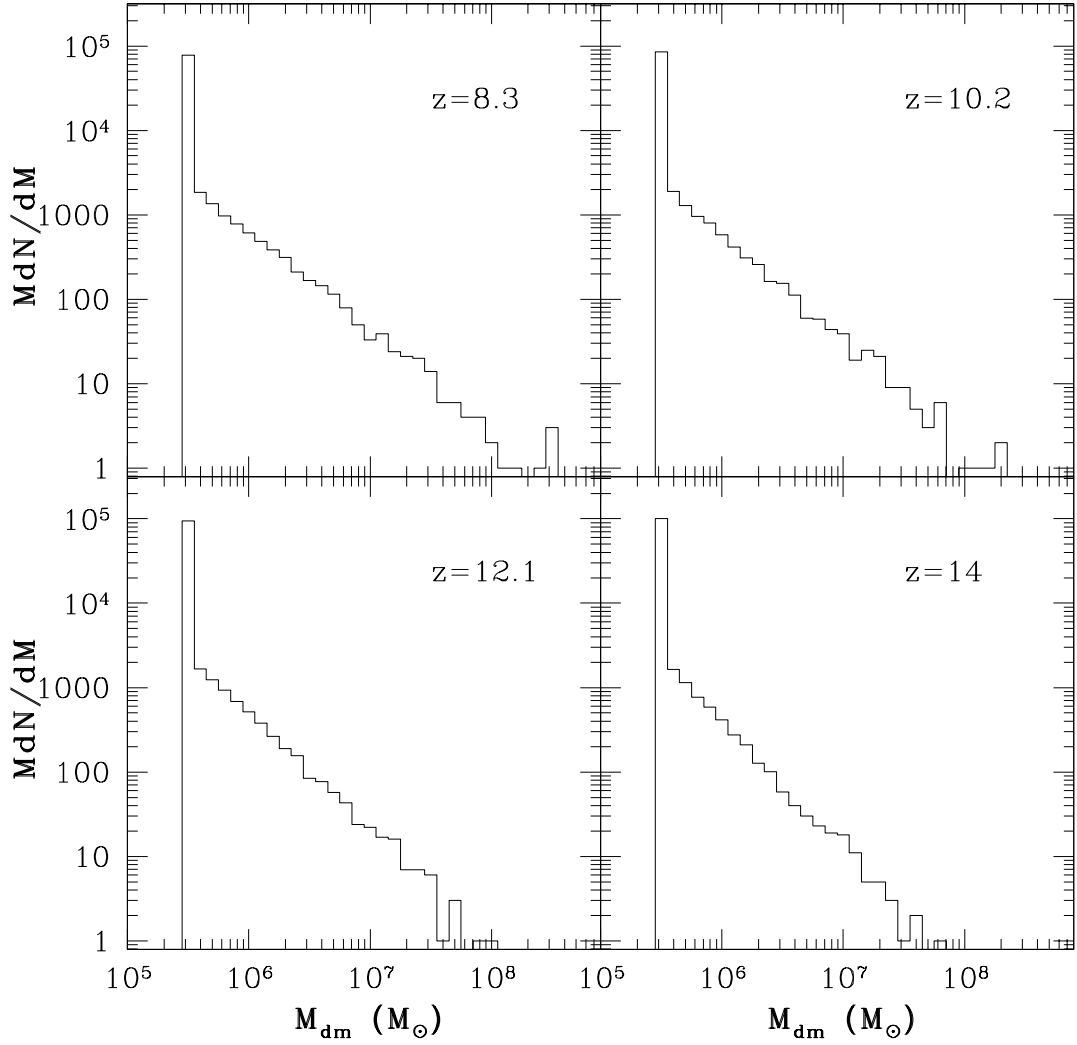}
\caption{Truncated mass function of the pre-reionization outputs from $z=8.3$ (top left), $z=10.2$ (top right), $z=12.1$ (bottom left), and $z=14$ (bottom right). The spike in the lowest mass bin at all four redshifts is due to the dark tracer particles.}
\label{FO.tmf}
\end{figure}

We turn the $1$~Mpc$^{3}$, $z=8.3$ output from the pre-reionization simulations to an equivalent cube of N-body particles as follows. First, any pre-reionization halo in the output becomes an N-body particle with a position, velocity, mass and, critically, unique ID. We then choose a mass resolution, $m_{trun}$, for our high resolution simulations and truncate the mass function of the pre-reionization halo at that resolution. To account for the additional mass needed to bring each $1$~Mpc$^{3}$ to the average density of the universe, we add dark, tracer particles. These tracer particles have a mass, $m_{trace}\simlt m_{trun}$. The positions and velocities of the tracer particles are determined from the position and velocities of the pre-reionization halos with $m<m_{trun}$. At the end of this process we have a $1$~Mpc$^{3}$ cube of N-body particles with $m_{min}\sim m_{trace}$, where positions, velocities and masses are determined by the $z=8.3$ output of the pre-reionization simulations. The mass function produced by this method is shown in the upper left panel of Figure~\ref{FO.tmf}.  The spike in the lowest mass bin shows the mass of the tracer particles. Critically, each particle has a unique ID allowing us to trace each pre-reionization halo to $z=0$ and retrieve its baryonic properties in the modern epoch.

We have generated a $1$~Mpc$^{3}$ cube for which each particle is a tracer for a pre-reionization halo. From the location of our Local Volume, we have a rectangular prism at $z=8.3$ where our high resolution region will go. We duplicate the $1$~Mpc$^{3}$ box to form the $m \times n \times p$ prism used for the high resolution region. We now have a $m \times n \times p$ prism with power on $l<$~Mpc scales. We add power on $l>$~Mpc scale using the position shift, $\delta {\bf{x}}$, of the coarse resolution particles via linear interpolation between them. Once we have $\delta {\bf{x}}$, we use the linear relation, $\delta {\bf{v}}=A(z)\delta {\bf{x}}$ to calculate the velocity perturbation, $\delta {\bf{v}}$, for our high resolution particles where $A(z)$ is the ratio of the $\delta {\bf{v}}/\delta {\bf{x}}$ at a given redshift. We now have a high resolution region  with $m \times n \times p$ embedded inside a $50^3$~Mpc$^{3}$ simulation at $z=8.3$. In the coarse resolution region, all power comes from the power spectrum, and in the high resolution region the power comes from the power spectrum on $l>$~Mpc scales and from the Press-Scheter via the pre-reionization simulations on $l<$~Mpc scales.

We then evolve the $50^3$~Mpc$^{3}$ volume from $z=8.3$ to $z=0$ using Gadget 2 \citep{Springel:05} with the outputs analyzed by AHF \citep{KnollmannKnebe:09}. These simulations are Runs A-C in Table 1.


\section{Second Order Initial Conditions}
\label{SO}

\begin{figure}
\centering
\includegraphics[height=120mm,width=120mm]{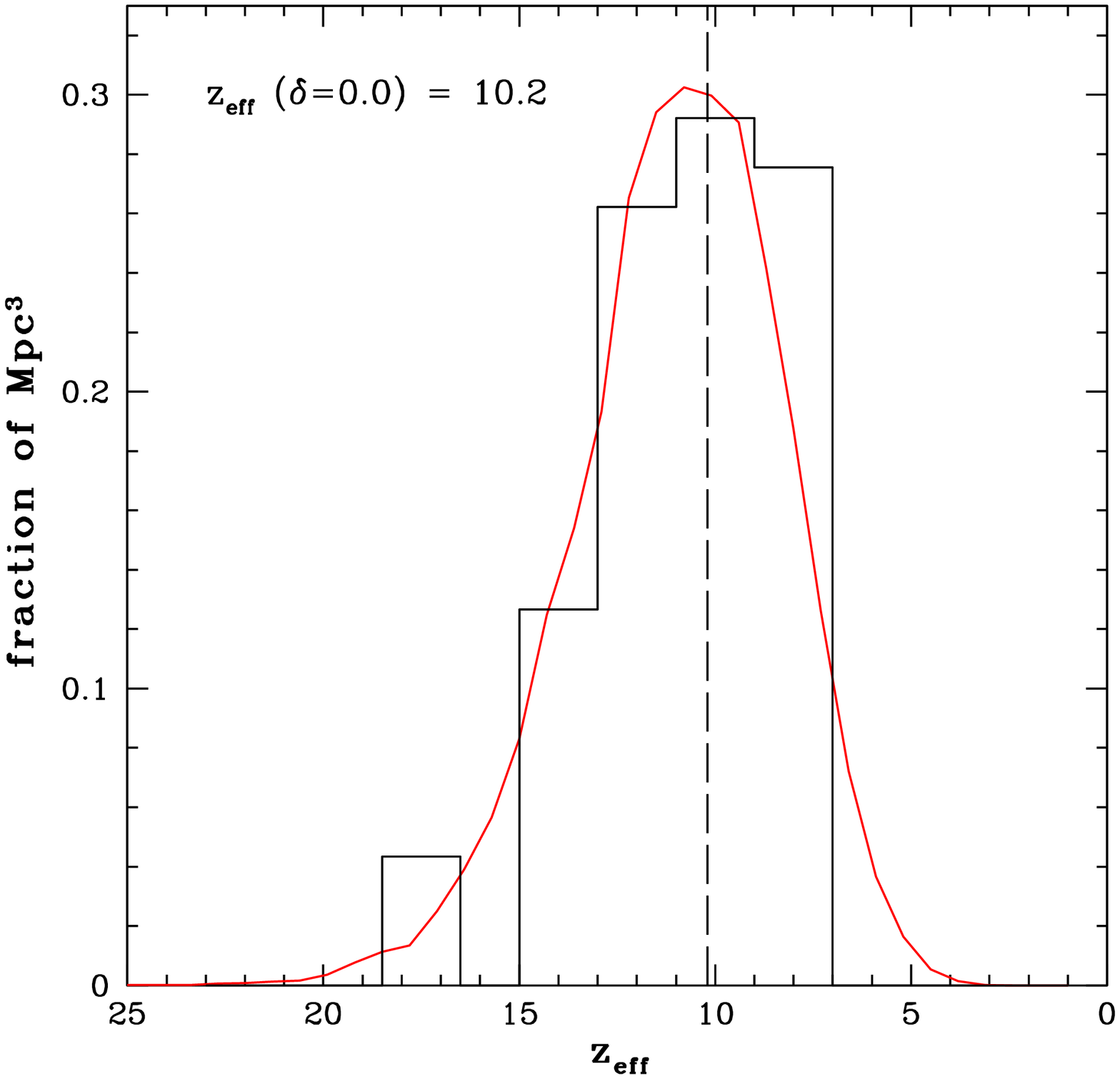}
\caption{Fraction of $1$~Mpc$^3$ cubes with a given effective redshift, $z_{eff}$.  The red curve shows the distribution for all the Mpc$^3$ cubes in our entire $50^3$~Mpc$^3$, low resolution volume. The black histogram show the fraction of $1$~Mpc$^3$ volumes within the high resolution region which use a given pre-reionization output.  Specifically, $z=(8.3,10.2,12,1,14)$. Sub-regions with $z_{eff}\sim17$ use the $z_{eff}=14$ pre-reionization output. The dashed vertical line shows the $z_{init}=10.2$ for our second order initial conditions.}
\label{SO.zeff}
\end{figure}

GK06 and our runs A-C assume the formation and evolution of all structures in the universe occur at the same rate. However, this is not valid. The rate at which structure forms and evolves is dependent on the local over-density or under-density of the universe, $\delta$ \citep{Cole:97}. The average density of the universe is $\delta=0.0$. Regions with $\delta < 0.0 $ form structures at a slower rate and vice versa for $\delta>0.0$. This difference affects the evolution of our primordial fossils in the following manner. Since dark matter halos will form at later times in the voids, they will accrete gas and begin to form stars at a lower $z$ compared to the filaments and clusters. Star formation is shut off in these minihalos by the reheating of the universe during reionization. If this occurs at the same time everywhere (X-ray reionization \citep{RicottiOstriker:04,RicottiOstrikerGnedin:05}) fewer of the minihalos in the voids will have formed stars, and those that did will have lower luminosities. 

Reionization begins in the clusters and filaments and spreads to the voids, reheating the lower density regions at lower redshift. Runs A-C would not need to account for this effect if the delay in halo and star formation in the voids exactly matched their later reheating. The second order initial conditions in Run D work in the model where the delay in reionizing and reheating in the voids is not enough to balance their later structure formation.

We begin the quantification of these different rates of evolution as a function of density, by assuming all parts of the universe evolve in a self-similar manner. Therefore, a $10^8 M_\odot$ halo in a cluster will collapse at a higher redshift than the same mass halo near a Milky Way, and that halo will form at a higher redshift than a $10^8 M_\odot$ halo in a void. We tie the $\delta$ to an effective redshift, $z_{eff}$:

\begin{equation}
\Delta z = (1+z_{init})((1+\delta)^{-0.6}-1),
\label{E.zeff}
\end{equation} 

where $z_{eff} = \Delta z + z_{init}$, with $z_{init}=z_{eff}(\delta=0.0)$, the current redshift of the entire universe, and $\Delta z$ is the shift due to the local under-density or over-density. Using this relation, we link the local density of the universe to the rate of structure formation via an effective redshift.  The distribution of $z_{eff}$ for our $50^3$~Mpc$^3$ volume is shown as the red curve in Figure~\ref{SO.zeff}. We now take the quantified relation between $z_{eff}$ and $\delta$, and apply it to our initial conditions. Before going any further, note that the selection of our Milky Ways and determination of our high resolution region are the same for Run D as for Runs A-C.

To account for the different rates of structure formation, we take additional steps when generating our high resolution region. (1) First, for each $1$~Mpc$^3$ cube within our coarse resolution volume, we calculate local density at $z=z_{init}$. (2) From the density, we use Equation~\ref{E.zeff} to find the shift in effective redshift due the over-density or under-density of each subvolume, and then calculate $z_{eff}$ (Figure~\ref{SO.zeff}). (3) Since we have a discrete set of pre-reionization outputs at $z=(8.3,10.2,12.1,14)$, we divide the $1$~Mpc$^3$ cubes within our high resolution region into four bins based on their densities and effective redshifts. The fraction of cubic Mpcs within our high resolution region in each effective redshift bin is shown as the black histogram overlaid on the $z_{eff}$ distribution in Figure~\ref{SO.zeff}. Note, that both the histogram and the smoother curve follow the same general shape. (4) Finally, based on which bin each Mpc$^3$ falls into, we assign it a pre-reionization output.

To account for the faster evolution in our high density regions, for Run D we use a $z_{init}=10.2$ instead of the $z_{init}=8.3$ used in Runs A-C. This allows us to assign a $z_{eff}=8.3$ to place near Milky Ways in our high density region. Nothing else substantially changes, except for using the $z=10.2$ output from the low resolution simulation to generate the $l>1$~Mpc structure in the high resolution region.

All the pre-reionization outputs are truncated to the same mass resolution using the same method described in the previous section.  As near as possible, we use tracer particles of the same mass in each pre-reionization output. The truncated mass functions of the three additional pre-reionization outputs used in our second order initial conditions are the additional panels (bottom and top right) in Figure~\ref{FO.tmf}.

After the high resolution has been built and the large scale modes added at $z_{init}=10.2$, we embed it inside the coarse resolution volume region and run to the present with Gadget 2.  Run D has two key differences when compared to Runs A-C.  First, the $50^3$~Mpc$^3$ snapshot used to generate the $l>1$~Mpc modes is $z=10.2$ instead of $z=8.3$. Second, the $z=8.3$ pre-reionization output is not used for the entire high resolution volume, but only in the over-dense regions, with outputs from $z=(10.2,12.1,14)$ used for the average and under-dense regions.

\bibliographystyle{../apj}
\bibliography{../../refs}

\end{document}